\documentclass[aps,article,nofootinbib,groupedaddress,twocolumn]{revtex4}
\usepackage{graphicx}
\usepackage{amsmath,amssymb}
\usepackage{hyperref}
\usepackage[usenames]{color}

\hypersetup{
    colorlinks=true,
    linkcolor=red,
    citecolor=blue,
}

\def\be{\begin{equation}}
\def\ee{\end{equation}}
\def\ba{\begin{eqnarray}}
\def\ea{\end{eqnarray}}
\def\nn{\nonumber}
\begin{document}

%\title{Cosmological tests of General Relativity using Principal Component Analysis}
%\title{Testing gravity on linear scales: a principal component analysis}
\title{Cosmological tests of General Relativity: a principal component analysis}

\date{\today}

\author{Alireza Hojjati$^{1}$, Gong-Bo Zhao$^{2}$,
Levon Pogosian$^{1}$, Alessandra Silvestri$^{3}$, Robert Crittenden$^{2}$, Kazuya Koyama$^{2}$}
\vspace{0.5cm}

\affiliation{
$^1$Department of Physics, Simon Fraser University, Burnaby, BC, V5A 1S6, Canada \\
$^2$Institute of Cosmology \& Gravitation, University of Portsmouth, Portsmouth, PO1 3FX, United Kingdom \\
$^3$Department of Physics, MIT, Cambridge, MA 02139, USA
}

\begin{abstract}
The next generation of weak lensing surveys will trace the evolution of matter perturbations and gravitational potentials from the matter dominated epoch until today. Along with constraining the dynamics of dark energy, they will probe the relations between matter overdensities, local curvature, and the Newtonian potential.  We work with two functions of time and scale to account for any modifications of these relations in the linear regime from those in the $\Lambda$CDM model. We perform a Principal Component Analysis (PCA) to find the eigenmodes and eigenvalues of these functions for surveys like DES and LSST. This paper builds on and significantly extends the PCA analysis of Zhao {\it et al.} (2009)~\cite{Zhao:2009fn} in several ways. In particular, we consider the impact of some of the systematic effects expected in weak lensing surveys. We also present the PCA in terms of other choices of the two functions needed to parameterize modified growth on linear scales, and discuss their merits. We analyze the degeneracy between the modified growth functions and other cosmological parameters, paying special attention to the effective equation of state $w(z)$. Finally, we demonstrate the utility of the PCA as an efficient data compression stage which enables one to easily derive constraints on parameters of specific models without recalculating Fisher matrices from scratch.

\end{abstract}

\maketitle

\section{Introduction}
The dynamics of linear scalar perturbations in alternative theories of gravity, as well as in exotic models of dark energy (DE) and dark matter, can differ significantly from that predicted by the cosmological concordance model, $\Lambda$CDM, even when the expansion histories are the same~\cite{Koyama:2005kd,Song:2006ej,Pogosian:2007sw,Dvali:2007kt,Silvestri:2009hh}. As in~\cite{Zhao:2009fn}, we will use the term {\it modified growth} (MG) when referring to all these models, including those based on General Relativity (GR).
It is expected that ongoing and upcoming tomographic weak lensing (WL) surveys, such as the {\it Dark Energy Survey}~(DES)\cite{DES} and {\it Large Synoptic Survey Telescope} (LSST)~\cite{LSST}, combined with the CMB and SNe data, will tightly constrain such modifications of growth dynamics on cosmological scales~\cite{Zhao:2009fn,Zhao:2008bn,Lombriser:2010mp,Song:2010fg, Zhao:2010dz,Giannantonio:2009gi,Daniel:2010ky,Bean:2010zq}.

Quite generally, a theory of gravity specifies how the metric perturbations relate to each other and how they are sourced by perturbations in the stress-energy tensor. In GR these relations are given by the {\it anisotropy} and {\it Poisson} equations, respectively. As in~\cite{Amendola:2007rr,BZ08,Zhao:2008bn,Zhao:2009fn,Pogosian:2010tj,Caldwell:2007cw,Jain:2007yk,arXiv:1109.4583}, we introduce two functions of time and scale, $\mu(a,k)$ and $\gamma(a,k)$, to allow for general departures of these equations from their $\Lambda$CDM form. By definition, these functions are equal to unity in $\Lambda$CDM, but generally have a time- and/or scale-dependence in alternative models of gravity and in models with clustering DE or a significant hot dark matter component, such as massive neutrinos.
Different, but in essence equivalent, parameterizations are used
 in~\cite{Hu:2007pj,BZ08,Amendola:2007rr,Daniel:2010ky,Bean:2010zq,Kunz:2006ca,Dore:2007jh,Acquaviva:2008qp,Daniel:2008et,Wei:2008vw}.

Such parametrizations can be used to test the validity of $\Lambda$CDM in a model-independent way, which is the main focus of this paper. If, instead, one aims to test a particular theory, there is no need to use these two functions, since one can derive the exact equations from the action and then calculate predictions for the observables to constrain the parameters (typically just a few) of the theory. However, as we will discuss in Section~\ref{sec:projection}, even in this case it can be advantageous to ``store'' information contained in observables into intermediate functions, such as our $\mu$ and $\gamma$. For instance, rather than modifying the standard Boltzmann codes, such as CAMB~\cite{camb,Lewis:1999bs}, differently for each specific model alternative to $\Lambda$CDM, one can modify them once to work for arbitrary $\mu$ and $\gamma$. Then, to evaluate observables in a given theory, it remains to work out its prediction for $\mu$ and $\gamma$. These may be approximate expressions, valid over a limited range of scales, or obtained by numerically solving a smaller set of equations to find $\mu$ and $\gamma$.

In~\cite{Zhao:2009fn}, a Fisher forecast and Principal Component Analysis (PCA) were performed to find the eigenmodes and eigenvalues of $\mu$ and $\gamma$ for surveys like DES and LSST, complemented with CMB and SNe data. The number of well-constrained modes approximately corresponds to the number of degrees of freedom (DoF) of these functions that can be measured by the surveys. Their scale and time dependence indicate the ranges where the surveys will be most likely to detect deviations from $\Lambda$CDM. The aim of this paper is to provide the details and to expand the study of \cite{Zhao:2009fn} in several ways. We present a detailed study of degeneracies between the MG functions and other cosmological parameters, paying special attention to the degeneracy with the DE equation of state $w(z)$. We also present the PCA of another choice of MG functions that helps to demonstrate the information content of WL surveys. We then investigate the effect of some of the systematics expected in WL surveys, and demonstrate the utility of the PCA approach as a data compression stage by using it to derive constraints on a parameter of a specific model.

\section{The formalism}
\label{sec:formalism}
\subsection{Evolution of linear perturbations}\label{sec:formalismA}
We consider linear scalar perturbations to the flat Friedmann-Robertson-Walker Universe, with the choice of the Newtonian gauge for the metric. The line element  reads
\begin{equation}\label{FRW}
ds^2 = a(\tau)^2[-(1+2\Psi)d\tau^2+(1-2\Phi)dx^2] \,,
\end{equation}
where $\Psi$ and $\Phi$ represent time- and space-dependent scalar perturbations of the metric and $\tau$ is the conformal time. We work in Fourier space, using the same symbols to indicate the perturbations in space and their Fourier transformed counterparts, i.e $\Psi=\Psi(a,k)$ and $\Phi=\Phi(a,k)$. The same convention is applied to the matter density contrast $\delta\equiv \delta\rho/\rho$ and the divergence of the velocity field $\theta \equiv i k^j v_j$. We assume that the matter perturbations obey the standard conservation equations, which for dark matter read:
\begin{eqnarray}
\label{matter-conservation}
\dot{\delta}+\theta -3\dot{\Phi}&=&0 \\
\dot{\theta}+ \mathcal{H} \theta - k^2 \Psi&=&0 \ ,
\label{matter-continuity}
\end{eqnarray}
where the dot denotes differentiation with respect to conformal time $\tau$, and $\mathcal{H} \equiv a^{-1}da/d\tau$. For the sake of simplicity we ignore radiation or baryonic effects, but they can be easily included if relevant.

One needs two additional equations to close the system for the four variables $\Phi$, $\Psi$, $\delta$ and $\theta$. These are normally provided by a theory of gravity, which prescribes how the two metric potentials relate to each other, and how they are sourced by the matter perturbations. 	Since we aim to test potential departures from $\Lambda$CDM, rather than working with a particular gravity theory, we close the system of equations by introducing two general functions of scale and time defined via:
\begin{eqnarray}\label{gamma}
\frac{\Phi}{\Psi} &\equiv &\gamma(a,k) \\
\label{parametrization-Poisson}
k^2\Psi &\equiv & - 4 \pi {a^2} G  \mu(a,k) \rho\Delta \,
\end{eqnarray}
where $\Delta \equiv \delta + 3\mathcal{H}\theta/k^2$ is the comoving matter density perturbation. Eqs.~(\ref{matter-conservation})-(\ref{parametrization-Poisson}) form a closed system that can be used to calculate the evolution of perturbations for any given functions $\mu$ and $\gamma$; they were extensively discussed in \cite{Pogosian:2010tj}. There, among other things, it was explicitly shown that they respect the superhorizon consistency conditions for adiabatic perturbations \cite{Bardeen:1980kt,Wands:2000dp} as long as $(k/\mathcal{H})^2 / (\mu \gamma) \rightarrow 0$ when $(k/\mathcal{H}) \rightarrow 0$.  In the Newtonian limit the functions $\mu$ and $\gamma$ are related to the Post-Newtonian-Parameters (PPN) of the Eddington-Robertson-Schiff formalism~\cite{will-book}. Specifically, $\mu\rightarrow\alpha_{{\rm PPN}}$, and $\gamma\rightarrow-\gamma_{{\rm PPN}}/\alpha_{{\rm PPN}}$  where $\alpha_{{\rm PPN}}$ and $\gamma_{{\rm PPN}}$ represent respectively the strength of gravity and the amount of curvature per unit mass.

By design, we have $\mu=\gamma=1$ in the standard cosmological model $\Lambda$CDM. Departures of $\mu$ and/or $\gamma$ from unity can happen if, for example, DE clusters or if it carries a non-negligible anisotropic stress. Alternatively, one could have $\mu \ne 1$ due to a significant fraction of massive neutrinos, which free stream on small scales. Finally, alternative gravity models generally predict scale- and time-dependent $\mu$ and/or $\gamma$ \cite{Silvestri:2009hh}.

As mentioned in Introduction, the main benefit of using these functions is that they allow for a model-independent test of  the growth dynamics on cosmological scales. Any measured deviation of either $\mu$ or $\gamma$ from unity would signal a departure from the $\Lambda$CDM model. It should be emphasized that $\mu$ and $\gamma$ do not necessarily have a simple form in specific models of MG, and generally depend on the choice of the initial conditions \cite{Skordis:2008vt,Ferreira:2010sz,Baker:2011jy}. For instance, in scalar-tensor models of gravity, the ratio of $\Phi$ and $\Psi$ is not a fixed function of $k$ and $a$. Instead, it is an expression that involves the time derivatives of $\Psi$ and $\Phi$. This means that $\mu$ and $\gamma$ correspond to {\it solutions} of equations of motion of a theory, rather than being a general prediction of a theory. Nevertheless, the functions $\mu$ and $\gamma$, while phenomenological in nature, are theoretically consistent and allow us to test for departures from $\Lambda$CDM independent of how complex the underlying theory of gravity is. One simply needs to be careful when translating the findings on $\mu$ and $\gamma$ into results on the parameters of specific models~\cite{Zuntz:2011aq}, paying attention to the choice of the initial conditions and possible additional simplifications, such as the quasi-static approximation.

Depending on the circumstances, such as the type of data available or the type of theory one wants to test, it can be more convenient to replace either $\mu$ or $\gamma$ with the function $\Sigma(a,k)$ defined as
\be\label{Sigma}
k^2(\Phi+\Psi) \equiv  - 8 \pi a^2 G \Sigma(a,k) \rho\Delta
\ee
The advantages of using different combinations of $\mu$, $\gamma$ and $\Sigma$ are discussed at length in~\cite{Pogosian:2010tj}. In this paper we will present and compare the results for $(\mu,\gamma)$ and $(\mu,\Sigma)$.

The implementation of this formalism in CAMB, which uses the synchronous gauge, is detailed in 
\cite{Zhao:2008bn,Hojjati:2011ix}.

\subsection{Principal Component Analysis}
\label{sec:formalismB}
Our goal is to determine how well $\mu(a,k)$ and $\gamma(a,k)$ can be constrained by future surveys, minimizing any assumption on the functions themselves. Therefore, rather than employing a specific expression for $\mu$ and $\gamma$, we will treat them as unknown functions of both time and scale, and bin them on a grid in the $(z,k)$ space (notice that we are now using the redshift $z$ as the time variable). With $m$ z-bins and $n$ k-bins,  we have $m\times n$ grid points to which we associate a value of the two functions. This is a $2\times 2$-dimensional problem, and in~\cite{Zhao:2009fn} we indicated these values with $\mu_{ij}$ and $\gamma_{ij}$. However, in practice, we transformed the 2D problem into a 1D one by mapping the grid into a chain, therefore transforming the matrices of values into two $m\times n$-dimensional vectors. We shall indicate the components of the vectors with $\mu_{i}$ and $\gamma_{i}$ where $i=1,\dots,m\times n$. In addition, we also bin the DE equation of state $w(z)$ in redshift, creating a $m$-dimensional vector and vary the usual cosmological parameters: the Hubble constant
$h$, cold dark matter density $\Omega_ch^2$, the baryon density
$\Omega_bh^2$, the optical depth $\tau$, the scalar spectral index
$n_s$, and the amplitude of scalar perturbations $A_s$. We assume that the bias is scale-independent on the linear scales considered in our analysis and introduce $N_b$ constant bias parameters, one for each photometric bin of the survey.

We then use the Fisher matrix formalism to estimate the anticipated covariance of our parameters $p_i$, $i=1,\dots,2\times m \times n + m + 6+ N_b$.
According to the Cramer-Rao theorem, any unbiased estimators for the parameters will give a covariance matrix that is not better than the inverse of the Fisher matrix of the parameters.
Therefore, after choosing our observables and experiments as described in next Section, we build the Fisher information matrix for the parameters $p_i$. Then, we invert it to determine the anticipated covariance matrix
\be
\label{covariance}
C_{ij} \equiv \langle (p_i-{\bar p_i})(p_j-{\bar p_j})\rangle \ ,
\ee
where ${\bar p_i}$ are the assumed best fit, or ``fiducial'', values.

Suppose that we want to know how well a given combination of experiments will measure $\mu$. We marginalize over the other parameters, and consider the $\mu$ block of the covariance matrix,  $C^{\mu}_{ij}$.
Since the individual pixels of $\mu$ are highly correlated, the covariance matrix will be non-diagonal, and the value of $\mu$ in any particular bin will be practically unconstrained. The PCA is a way to decorrelate the parameters and find their linear combinations that are best constrained by data. Namely, we solve an eigenvalue problem to find a matrix $W^{\mu}$ that diagonalizes $C^{\mu}$:
\be
C^{\mu} = (W^{\mu})^T \Lambda W^{\mu} \ ;  \ \ \Lambda_{ij} = \lambda_i \delta_{ij} \ ,
\label{rotate}
\ee
where $\lambda_i$'s are the eigenvalues. Smaller values of $\lambda_i$ correspond to the better constrained linear combinations of $\mu$'s:
\be
\alpha_i = \sum_{j=1}^{m \times n} W^{\mu}_{ij} (\mu_j-{\bar \mu_j})  \ .
\label{alphas}
\ee
One can think of $\alpha$'s as the new set of uncorrelated parameters obtained by a rotation of $\mu$'s, with the error on $\alpha_i$ given by $\sqrt{\lambda_i}$. In practice, one finds that only a few of the $\alpha$'s are well constrained (i.~e. their $\lambda$'s are small), while most are essentially unconstrained. This is the main benefit of performing a PCA -- it takes a function with many (infinite) degrees of freedom and isolates their few uncorrelated linear combinations that can be constrained by a given experiment. By construction, $W^TW=I$, so Eq.~(\ref{alphas}) can be inverted as
\be
\mu_i-{\bar \mu_i} = \sum_{j=1}^{m \times n} W^{\mu}_{ij} \alpha_j \ .
\label{inverted1}
\ee
where $i$ labels a point on the $(z,k)$ grid and $j$ label the eigenmode. Thus, taking the continuous limit, and using $\bar{\mu}=1$ as the fiducial value, we can formally rewrite this as
\be
\mu(z,k) = 1+ \sum_j \alpha_j  W^{\mu}_{j} (z,k) \ ,
\label{inverted2}
\ee
which is an expansion of $\mu$ into an orthogonal basis of eigenvectors $W^{\mu}_{j} (z,k)$. We can now rearrange the eigenvectors into a 2D form and plot them as surfaces in the $(z,k)$ space. We will refer to these surfaces as the principal components (PC's) or {\it eigenmodes}; the shapes of the best constrained eigenmodes indicate the kind of features in $\mu$ that experiments are most likely to constrain. The regions in $(z,k)$ where the best constrained eigenmodes peak indicate the {\it sweet spots}, i.e. the intervals in time and scale where a given experiment will be more sensitive to departures from $\Lambda$CDM. The number of nodes in the $z$ and $k$ directions indicate the degree of sensitivity of the surveys to the $z$- and $k$-dependence of $\mu$. The same procedure can be repeated for the function $\gamma(a,k)$ (or $w(a)$), in this case isolating and diagonalizing the $\gamma$ block of the covariance matrix to determine the eigenvectors $W^{\gamma}(z,k)$ and the corresponding eigenvalues.

The procedure outlined above addresses the ability of data to constrain $\mu$ and $\gamma$ separately\footnote{We note that in this paper we do not attempt to reconstruct  $\mu(a,k)$ and $\gamma(a,k)$ from data, nor we propose PCA as a reconstruction tool. Instead we forecast the ability of surveys to detect departures of $\mu$ and $\gamma$ from unity and use PCA to determine the best constrained eigenmodes.}.  Namely, when deriving the eigenmodes and eignevalues of $\mu$ ($\gamma$) we marginalize over $\gamma$ ($\mu$).  However, observations probe combinations of the potentials $\Phi$ and $\Psi$ that depend on both $\mu$ and $\gamma$. This yields a degeneracy between $\mu$ and $\gamma$ and,  by marginalizing over one, we lose the information that is common to both functions. In addition to forecasting separate constraints on $\mu$ and $\gamma$, one may want to know how sensitive data is to \emph{any} departure from the standard growth. Namely, we may want to answer a less ambitious, yet equally interesting, question of whether either of the two functions deviates from unity, without specifying which. For this purpose, we want to save the information common to both functions, which we previously lost by mutual marginalization. Hence, we consider the combined principal components of $\mu$ and $\gamma$. We follow the same procedure as before, except now we diagonalize the block of the covariance matrix containing the pixels of $\mu$ \emph{and} $\gamma$. The components of the matrix that diagonalizes $C^{\mu\gamma}$ will be $W^{\mu\gamma}_{ij}$ where now $i,j=1....2\,m\times n$; each eigenmode $j$ consists now of a double series of pixels on the $(k,z)$ grid, representing combined eigenmodes in the two sub-spaces.

\subsection{Covariance matrix for $\Sigma$}\label{sec:formalismC}

As discussed in~\cite{Pogosian:2010tj}, the choice of two functions parametrizing MG is not unique. Depending on the circumstances, it can be preferable to replace $\gamma$ with the function $\Sigma$ defined in Eq.~(\ref{Sigma}), and work with the combination ($\mu$,$\Sigma$). In that case, one way to proceed is to repeat the procedure outlined in Sec.~\ref{sec:formalismB} for the new combination ($\mu$,$\Sigma$). Alternatively, one can use the information already stored in the ($\mu$,$\gamma$) pixels to infer the covariance matrix for $\Sigma$, which is what we proceed to do. From Eqs.~(\ref{gamma})-(\ref{Sigma}), we have
\begin{equation}
\Sigma= \frac{1}{2} \mu (1+\gamma)\,.
\end{equation}
Then, pixelizing $\Sigma$ on the same $(k,z)$ grid, we can derive its covariance matrix in terms of the covariance matrix elements for $\mu$ and $\gamma$ as
\begin{eqnarray}
 C^{\Sigma\Sigma}_{ij} &=& \frac{1}{4} [ \mu_i\mu_j  C^{\gamma\gamma}_{ij} + (1+\gamma_i) (1+\gamma_j) C^{\mu\mu}_{ij} \nn \\
&+& \mu_i(1+\gamma_j) C^{\gamma \mu}_{ij}+\mu_j(1+\gamma_i)C^{\mu\gamma }_{ij} ],
\end{eqnarray}
where, for example, $C^{\gamma \mu}_{ij}$ is the covariance between $\gamma_i$ and $\mu_j$. Analogously, one can derive the covariance of $\Sigma$ with $\mu$, $C^{\Sigma\mu}$.

\section{The observables}
\label{sec:observables}

The ongoing and future tomographic large scale structure surveys (like DES~\cite{DES}, PAN-STARR~\cite{PAN-STARR} and LSST~\cite{LSST}) will provide measurements of galaxy number counts (GC) and weak lensing (WL) on the same patch of sky, as well as a large number of supernovae (SN). This, in combination with the full sky CMB data from WMAP and Planck, allows us to consider all possible cross-correlations: CMB-WL, CMB-LSS, LSS-WL, in addition to the three auto-correlation functions. This information can be used to forecast constraints on the differences between the metric potentials and the scale-time variation of the effective Newton constant, parametrized respectively by $\gamma$~(\ref{gamma}) and $\mu$~(\ref{parametrization-Poisson}). In what follows, we provide a brief overview of the observables and the assumptions about the experiments which only have minor differences from the assumptions used in \cite{Zhao:2008bn}.

\subsection{Angular spectra}

For any two fields, $X(\mathbf{\hat{n}})$ and $Y(\mathbf{\hat{n}})$, measured by an observer as function of the direction on the sky $\mathbf{\hat{n}}$, one can define the angular power spectrum $C_\ell^{XY}$ via
\begin{equation}
C^{XY}(\theta) = \sum_{\ell=0}^{\infty} \frac{2\ell+1}{4 \pi}C_\ell^{XY} P_\ell(\cos\theta) \ ,
\label{eq:leg_series}
\end{equation}
where $C^{XY}(\theta) \equiv C^{XY}(|\mathbf{\hat{n}}_1 - \mathbf{\hat{n}}_2|) \equiv \left<X(\mathbf{\hat{n}}_1)Y(\mathbf{\hat{n}}_2)\right>$ is the two-point correlation function, and $P_\ell$ are the Legendre functions. In a flat universe, $C_\ell^{XY}$ can be expressed in terms of the primordial curvature power spectrum $\Delta_{\cal R}^{2}$ and the angular transfer functions $I_{\ell}^{X,Y}(k)$ as
\begin{equation}
C_\ell^{XY}= 4\pi \int \frac{dk}{k} \Delta_{\cal R}^{2}
I_{\ell}^X(k) I_{\ell}^Y(k), \label{eq:gen}
\end{equation}
where $I_{\ell}$ are the transfer functions defined as
\begin{equation}
I_{\ell}^X(k) =  \int_0^{z_*} d z W_X(z)  j_\ell[kr(z)]\tilde{\cal X}(k, z).
\label{eq:I_gen}
\end{equation}
and similarly for $I_\ell^Y$. 
In the above, $z_*$ is a sufficiently high redshift at which the initial condition for the mode $k$ is specified, $j_\ell$ are the spherical Bessel functions, $r(z)$ the comoving distance to a point at redshift $z$, and $W_X$ are the window functions which, depending on the observable, specify the range of redshifts contributing to $X$. Finally, $\tilde{\cal X}(k, z)$ is the Fourier transform of the three-dimensional field ${\cal X}(\hat{n}r(z),z)$ responsible for producing the two-dimensional observable $X$ ({\it i.e.} $X(\hat{n})=\int_0^\infty dz \,W_X(z){\cal X}(\hat{n}r(z),z)$\,).
A detailed derivation of the above expressions is given in \cite{Zhao:2008bn}. We adopt adiabatic initial conditions as detailed in \cite{Ma:1995ey,Garriga:2003nm}. The observable quantities for which we evaluate $I_\ell$'s are GC in several photometric redshift bins, the WL shear in several bins, and the CMB temperature anisotropy.

For GC, we have
\begin{equation}
I_{\ell}^{G_i}(k) =  b_i \int_0^{z_*} d z W_{G_i}(z)  j_\ell[kr(z)]\delta(k, z) \ ,
\label{eq:I_delta}
\end{equation}
where $b_i$ is the bias, $W_{G_i}(z)$ is the normalized selection function for the $i$th redshift bin, and $\delta(k, z)$ is the density contrast transfer function (and we have dropped the tilde following the convention outlined in Sec.~\ref{sec:formalismA}). We work under the assumption that on large scales the bias can be treated as scale-independent and can be modeled with one free parameter $b_i$ for each redshift bin $i$.

For weak lensing, the relevant $I_\ell$'s are given by
\be
\label{wls}
I_{l}^{\kappa_{i}}(k)=\int_0^{z_*} dz
W_{\kappa_{i}}(z) j_{l}[kr(z)] (\Psi+\Phi) \ ,
\ee
where $W_{\kappa_{i}}(z)$ is the window
function for the $i$th  bin of sheared galaxies with a normalized
redshift distribution $W_{S_i}(z)$, {\it i.e.}:
\be
W_{\kappa_{i}}(z)=\int_z^{\infty} dz' \frac{r(z')-r(z)}{r(z)}
W_{S_i}(z') \ .
\ee

The transfer functions for the CMB temperature anisotropy receive contributions from the last-scattering surface (at $z \sim 1100$) and from more recent redshifts via the Integrated Sachs-Wolfe (ISW) effect. The modifications of gravity considered in this paper are negligible at recombination. Therefore, their only imprint on the CMB will be via the ISW effect. For the ISW contribution to the CMB, we have
\begin{equation}
I_{\ell}^{ISW}(k) =   \int_0^{z_*} dz e^{-\tau(z)} j_{\ell}[kr(z)] {\partial \over \partial z}\left[\Psi+\Phi\right] \ ,
\label{eq:I_ISW}
\end{equation}
where $\tau(z)$ is the opaqueness function.

We numerically evaluate the transfer functions $I_\ell$ using the first version of the publicly available code MGCAMB ({\it Modified Growth with CAMB})~\cite{Zhao:2008bn}\footnote{A new version of MGCAMB was recently introduced in \cite{Hojjati:2011ix}, and is publicly available at \url{http://www.sfu.ca/~aha25/MGCAMB.html}. The first version was based on CAMB-Sources, which made it easy to evaluate WL and GC spectra, but was not compatible with CosmoMC. The later version is written as a patch for CAMB, is compatible with CosmoMC, but does not evaluate WL and GC spectra yet.} and obtain the angular spectra
$C_{l}^{XY}$. A joint analysis of CMB and data from a tomographic lensing survey with $M$ GC redshift bins and
$N$ WL bins gives us a total of $3+M(M+1)/2+N(N+1)/2+M+N+MN$
different types of $C_{\ell}$'s respectively from CMB, GC, WL, GC$\times$CMB,
WL$\times$CMB and WL$\times$GC, (we do not correlate
CMB polarization with GC and WL). For example, combining Planck with
DES, with $M=4$ GC bins and $N=4$ WL bins, gives us $47$ different
types of spectra. A combination of Planck with LSST, with $10$ GC
bins and $6$ WL bins, gives us $155$ different $C_{\ell}$'s. 

We only use the parts of the spectra that correspond to the linear
cosmological regime. Including higher $\ell$, or smaller scales,
would require us to account for non-linear effects which, strictly
speaking, are not allowed within our framework. To accurately model
growth on non-linear scales, one needs input from N-body
simulations, which can only be performed for specific modified
gravity theories. The fact that we are not testing a specific model,
but constraining a general departure from GR, defined in terms of
linear perturbation variables, precludes us from having a reliable
description of non-linear corrections. We restrict ourselves
to the linear regime by cutting off the $C_{\ell}^{XY}$ spectra at
$\ell_{\rm max} \sim 0.2~h\,r(z_{s})$. This cutoff
roughly corresponds to $k_{\rm max}\sim 0.2\,h {\rm Mpc}^{-1}$ at
$z=0$. There is certainly a wealth of information about MG parameters on smaller scales \cite{Beynon:2009yd}, 
and while it would be tempting to include it in our analysis, it would make our predictions unreliable since our analysis is limited to linear theory.

\subsection{Fisher matrices}
\label{sec:fisher}

To determine how well the surveys will be able to constrain
our model parameters, we employ the standard Fisher matrix
technique~\cite{Fisher}. The inverse of the Fisher matrix $F_{ab}$
provides a lower bound on the covariance matrix of the model
parameters via the Cram$\acute{\rm e}$r-Rao inequality, ${\bf C}
\geq {\bf F}^{-1}$~\cite{Fisher}. For zero-mean Gaussian-distributed
observables, like the angular correlations $C^{XY}_\ell$, the Fisher matrix is given by
\be
F_{ab} =
f_{\rm sky} \sum_{\ell=\ell_{\rm min}}^{\ell_{\rm max}}\frac{2\ell +
1}{2} {\rm Tr}\left( \frac{\partial {\bf C_\ell}}{\partial p_a} {\bf
\tilde{C}_\ell^{-1}}\frac{\partial {\bf C_\ell}}{\partial p_b} {\bf
\tilde{C}_\ell^{-1}} \right) \ ,
\label{eq:Fisher}
\ee
where $p_{a}$ is the ${a}^{\rm th}$ parameter of our model and ${\bf
\tilde{C}_\ell}$ is the ``observed'' covariance matrix with elements
$\tilde{C}^{XY}_\ell$ that include contributions from noise:
\be
\tilde{C}^{XY}_\ell= C^{XY}_\ell+N^{XY}_\ell \ .
\label{eq:NoiseAdd}
\ee
 Eq.~(\ref{eq:Fisher}) assumes that all fields
$X(\hat{\bf n})$ are measured over contiguous regions covering a
fraction $f_{\rm sky}$ of the sky. The value of the lowest multipole
can be approximately inferred from $\ell_{\rm min} \approx \pi
/(2f_{\rm sky})$. It is also possible to write expressions for
separate contributions to the Fisher matrix from particular subsets
of observables, as detailed in \cite{Zhao:2008bn}.

The noise matrix $N^{XY}_\ell$ includes the statistical noise as well as the expected
systematic errors. Systematics are notoriously difficult
to predict, and are often ignored in forecasts.
In this paper, we consider two cases -- the case when $N^{XY}_\ell$ includes the statistical noise only, and the case
when certain types of systematics are included. We follow \cite{Huterer:2005ez,Zhan:2008jh} and consider three sources of systematics for future tomographic surveys: photo-$z$ errors, as well as additive and multiplicative errors due to the uncertainty of the point spread function (PSF) measurements. The details of our modeling of the systematic effects are presented in Appendix~\ref{sec:appendix}. Our assumptions about the statistical errors in CMB, CG and WL are presented in the following subsection.

For supernovae, the information matrix is
\be
F^{\mathrm{SN}}_{ab}= \sum_{i}^{N} {1\over
\sigma(z_i)^2 } {\partial m(z_i)\over \partial p_a} {\partial
m(z_i)\over
\partial p_b}.
\label{fisher:sne}
\ee
where $m(z)$ is their redshift-dependent
magnitude, the summation is
over the redshift bins, and $\sigma(z_i)=0.13$ (see Sec. \ref{sec:cmbsn} for details).

Given a set of theoretical covariance matrices over a given
multipole range, and the specifications for the expected noise in
specific experiments, we can compute the Fisher matrix. The
derivatives with respect to the parameters (pixels of $\mu$ and $\gamma$ in our case) are computed numerically using finite differences.

\subsection{Experiments}
\label{sec:experiments}

The data considered in our forecasts include CMB temperature and polarization (T and E), WL of distant galaxies, GC, their
cross-correlations, and SNe observations. We assume CMB T and
E data from the Planck satellite~\cite{Planck}, the galaxy
catalogues and WL data by the \emph{Dark Energy Survey}~\cite{DES} and \emph{Large Synoptic Survey Telescope} (LSST)~\cite{LSST}, complemented by a SNe data set provided by the futuristic Euclid-like survey \cite{euclidsn}.
\subsubsection{DES and LSST}

We take the total galaxy number density to be given by
\begin{equation}
N_G(z) \propto z^2 {\rm exp}(-(z/z_0)^2) \ ,
\end{equation}
which is a slight modification of the model due to Wittman {\it et al.}~\cite{Wittman}. The parameter $z_0$ depends on the experiment and defines the redshift at which the most galaxies will be observed. The galaxies can be divided into photometric redshift bins, labeled with index $i$,
\begin{equation}
N_G(z) = \sum_i N_{G_i}(z).
\end{equation}
In our analysis, we assume that the photometric
redshift errors are Gaussian distributed,
and that their rms fluctuations increase with redshift as
$\sigma(z)=\sigma_\mathrm{max}(1+z)/(1+z_\mathrm{max})$.
The bin sizes are chosen to increase proportionally to the errors.
The resulting photometric redshift distributions are given by
\begin{equation}
N_{G_i}(z) = \frac{1}{2} N_G(z)\left[
  \mathrm{erfc}\biggl( \frac{z_{i-1}-z} {\sqrt{2} \sigma(z)}\biggr)
- \mathrm{erfc}\biggl( \frac{z_i-z}{\sqrt{2} \sigma(z)}\biggr) \right],
\label{eq:erfc}
\end{equation}
where erfc is the complementary error function. For a given photometric redshift bin, the normalized window function that appears in Eq.~(\ref{eq:I_delta}) is given by
\begin{equation} \label{windowfunction}
 W_{G_i}(z) = \frac{N_{G_i}(z)}{N^i} \,
\end{equation}
where $N^i$ is the total number of galaxies in the $i$th bin.

DES is a project aimed at studying the
nature of the cosmic acceleration, and is planned to start
observations in 2012~\cite{DES}. DES includes a 5000
square degree multi-band, optical survey probing the redshift range
$0.1 < z < 1.3$ with a median redshift of $z=0.7$ and an approximate
1-$\sigma$ error of $0.05$ in photometric redshift. In our
simulation, for both WL and GC, we assume a sky fraction $f_{\rm{sky}}=0.13$, and an angular density of galaxies
$N_G=10$ gal$/$arcmin$^2$. We also assume $\gamma_{\rm rms}=0.18+0.042\,z$, which is the rms shear stemming from the intrinsic
ellipticity of the galaxies and measurement noise, and the photometric redshift uncertainty given by $\sigma(z)=0.05(1+z)$.

LSST is a proposed large
aperture, ground-based, wide field survey telescope~\cite{LSST}. It
is expected to cover up to half of the sky and
catalogue several billion galaxies out to redshift $z\sim 3$. For LSST forecasts, we adopt parameters from the recent review paper by the LSST
collaboration~\cite{Ivezic:2008fe}. Namely, we use $f_{\rm{sky}}=0.5$, $N_{G}=50$ gal$/$arcmin$^2$ for both WL and counts, $\gamma_{\rm
rms}=0.18+0.042\, z$, and $\sigma(z)=0.03\,(1+z)$.

For both DES and LSST, we take the GC photometric bins to be
separated by $5\sigma(z)$. This leads to four redshift bins for DES
and ten for LSST. For WL (source) galaxies, we use four bins for DES
and six for LSST.

\subsubsection{CMB, Supernovae, and other priors on cosmological parameters}
\label{sec:cmbsn}

In our forecasts, we assume spatially flat geometry and parametrize the dark energy equation of state using 20 bins from $z=0$ to $z=3$ uniform in $z$, and one wide bin from $z=3$ to $z=1100$.  In addition to the MG parameters, and the $N_b \equiv M$ bias parameters, we vary $h$, $\Omega_ch^2$, 
$\Omega_bh^2$, $\tau$, $n_s$, and $A_s$. Their
fiducial values are taken to be those from the WMAP 7-year data best
fit~\cite{wmap7}: $\Omega_b h^2 = 0.023, \Omega_c h^2 =
0.11, h=0.71, \tau = 0.088, n_s=0.963$. The fiducial values for
bias parameters are motivated by the parametrized halo model
described in \cite{HuJain}. Imposing a prior on the value of
$h$ from the Hubble Space Telescope (HST) did not make a noticeable
difference in our results.

We assume the expected CMB data from the Planck mission~\cite{Planck} of the \emph{European Space Agency} (ESA) using the same parameters as in \cite{Zhao:2008bn}. In addition, to better constrain the background expansion parameters, we include simulated SNe luminosity data for a Euclid-like survey, {\it e.g.}, the one proposed in \cite{euclidsn}.
We generate 4012 data points randomly distributed
in 14 redshift bins from $z = 0.15$ to $z = 1.55$, and combine
300 low-z SNe from the Nearby Supernova Factory (NSNF) survey \cite{NSNF}. We calculate the exact distance modulus for each
model, and put a Gaussian noise with rms error of $\sigma=0.13$
to displace all the data points. The absolute magnitude, or the so-called nuisance parameter ${\cal M}$, is treated as an undetermined parameter in our analysis.

Note that DES and LSST will produce their own SNe luminosity measurements that we did not include in our forecasts. Instead, we assume that a high quality SNe catalogue will eventually become available, and focus on the dependence of MG constraints on the quality of
the WL and GC data.

\section{The PCA of linearized MG}
\label{sec:mgpca}
In this Section we analyze the principal components (eigenmodes) of the functions $\mu$ and $\gamma$  defined in Sec.~\ref{sec:formalism} for the combination of large scale structure (WL and GC), CMB and SNe experiments detailed in Sec.~\ref{sec:experiments}. In particular, we want to investigate the effect of degeneracies with other cosmological parameters on the eigenmodes and eigenvalues of $\mu$ and $\gamma$. For this purpose, in Subsection~\ref{pca-all-fixed}, we start with the simplest case where only uncertainties in the pixelated functions  $\mu$ and $\gamma$ (referred to as ``MG pixels'') are considered, with all other parameters fixed to their fiducial values. The highest redshift pixels ($3<z<30$) of $\mu$ and $\gamma$ are outside the range directly probed by the WL surveys. However, they do impact the observables: $\mu(z>3,k)$ re-sets the amplitude of the growth at all lower redshifts, while $\gamma(z>3,k)$ affects the ISW contribution to the CMB temperature anisotropy. The sensitivity of observables to variations in $\mu(z>3,k)$ and $\gamma(z>3,k)$ depends on the assumed high-$z$ cutoff -- making the high-z interval wider increases the sensitivity. Thus below we focus on quantities that are independent of this cutoff, such as the eigenmodes and eigenvalues of $\mu$ and $\gamma$ obtained after marginalizing over the high-z MG pixels, as well as the uncertainty in the overall growth between recombination and $z=3$, which is directly controlled by $\mu(z>3,k)$, but is independent of the choice of the high-$z$ cutoff (see \ref{error-r}). In \ref{MObias}, we discuss effects of marginalizing over the high-$z$ pixels and galaxy bias parameters. In (\ref{MOeverything}), we consider degeneracies with the cosmological vanilla parameters $\{\Omega_b{h}^{2},\Omega_c{h}^{2},h, \tau, n_s,A_s\}$ and the effective equation of state $w(z)$. Throughout this Section, we always marginalize over the SNe nuisance parameter. To gain additional insight, we also project our findings onto the function $\Sigma$.

We examine the parameter degeneracies using LSST as the fiducial survey for WL and GC. Then, in \ref{des-compare}, we compare the final case, in which we marginalize over all the parameters except the $\mu$ and $\gamma$ in the range $0<z<3$, to the analogous forecast for DES.

\subsection{PCA for LSST with all parameters, except MG pixels, fixed (high-z information included)}
\label{pca-all-fixed}

\begin{figure}[tbp]
\includegraphics[width=1.05\columnwidth]{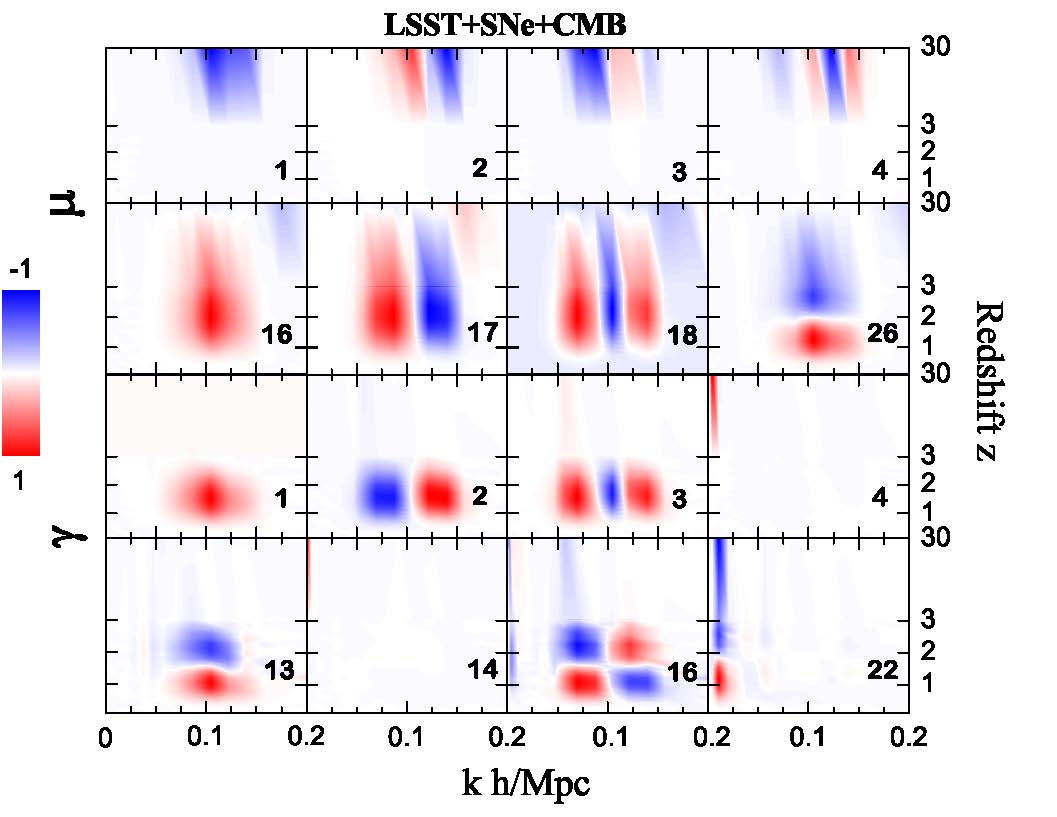}
\caption{The eigenmodes of $\mu$ and $\gamma$ for LSST(+SN+CMB) with all other parameters fixed to fiducial values.}
\label{fig:IV_A_mugamma}
\end{figure}

\begin{figure}[tbp]
\includegraphics[width=1.05\columnwidth]{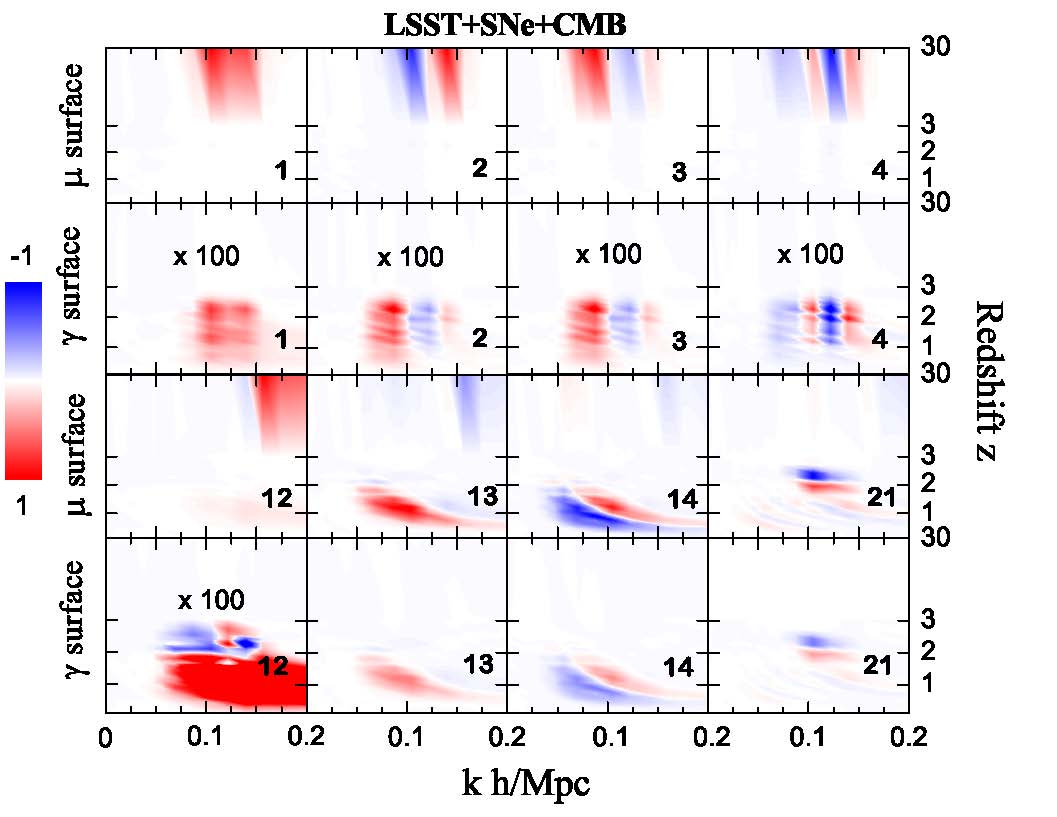}
\caption{The {\it combined} eigenmodes of $(\mu,\gamma)$ for LSST(+SN+CMB) with all other parameters fixed to fiducial values.}
\label{fig:IV_A_combined}
\end{figure}

\begin{figure}[tbph]
\includegraphics[width=1.05\columnwidth]{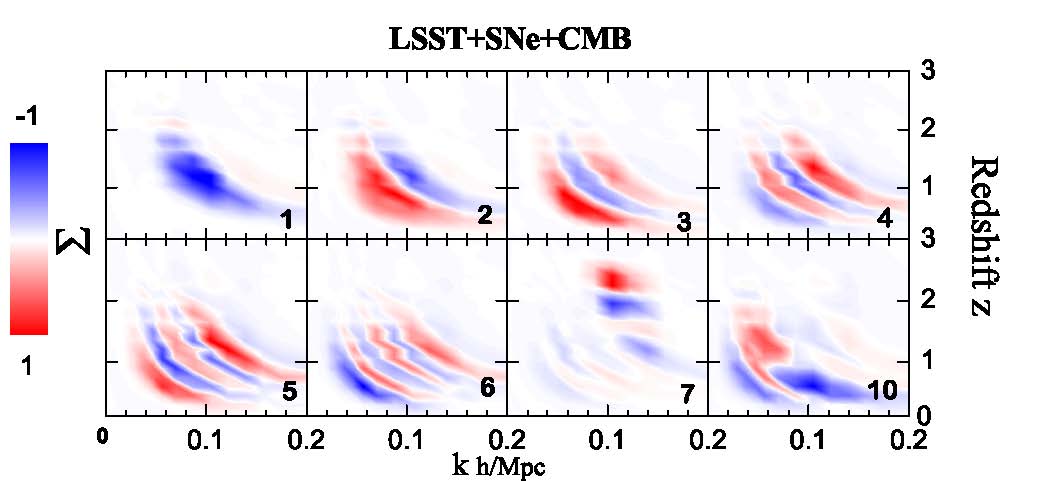}
\caption{The eigenmodes of $\Sigma$ for LSST(+SN+CMB) with all other parameters fixed to fiducial values.}
\label{fig:IV_A_Sigma}
\end{figure}

\begin{figure}[tbph]
\includegraphics[width=1.05\columnwidth]{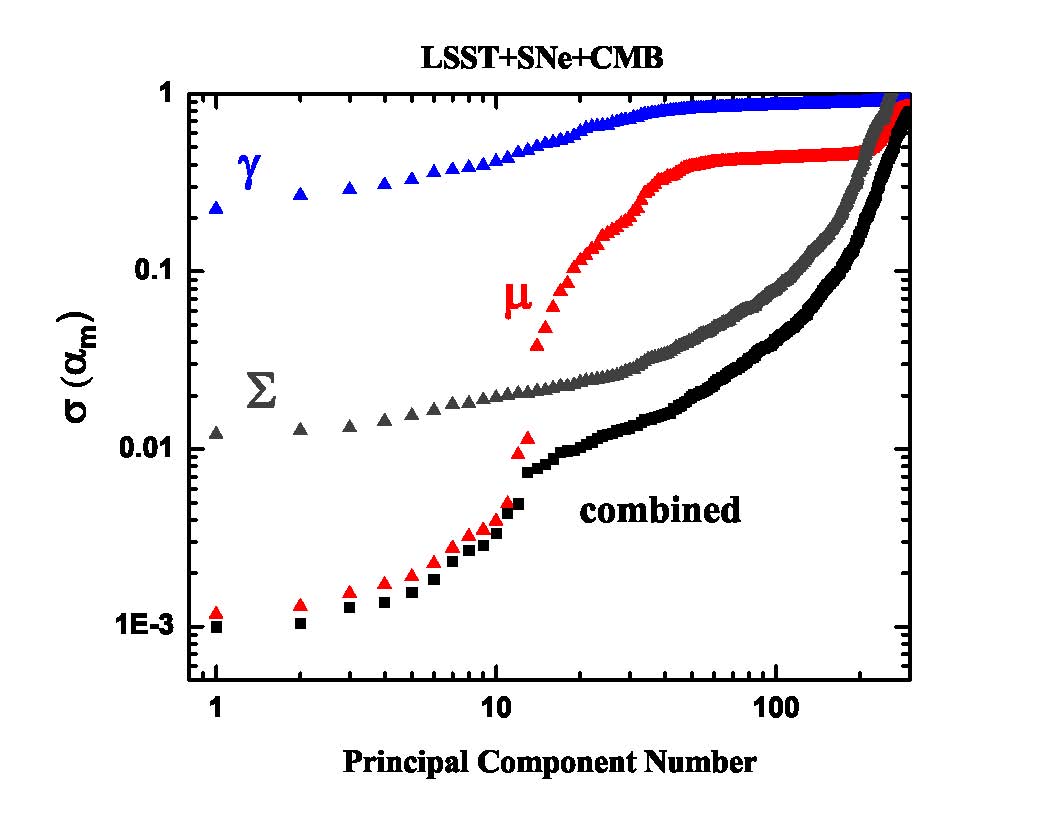}
\caption{The uncertainties (square roots of covariance eignevalues) associated with the eigenmodes of $\mu$, $\gamma$, $\Sigma$ and the combined $(\mu,
\gamma)$ case for LSST(+SN+CMB) with all other parameters fixed to fiducial values.}
\label{fig:IV_A_errors}
\end{figure}

As a first step, we study the eigenmodes and associated uncertainties of the MG functions without considering their covariance with other parameters. We do this by isolating and inverting the block of the Fisher matrix containing only MG pixels (including the high-z bins) and diagonalizing the resulting covariance matrix.

Fig.~\ref{fig:IV_A_mugamma} shows some of the eigenmodes of $\mu$ and $\gamma$. Each panel in these plots represents a region in $(z,k)$ space with an eigenmode function plotted as a surface, as described in Sec.~\ref{sec:formalismA}. We order, and consequently number, the eigenmodes according to how well they can be constrained, {\it i.e.} following the ordering of the corresponding errors (square root of the covariance matrix eigenvalues) from the smallest (best constrained) to the largest (least constrained).

The first feature to notice is in the $\mu$ eigenmodes (top two rows in Fig.~\ref{fig:IV_A_mugamma}), where the best constrained modes peak at high-$z$, and show no features at low z;  we need to get to the 16th mode to start seeing some features in the $z<3$ interval, which is the actual redshift range of LSST. This is because $\mu$ directly affects the growth of matter density perturbations, and changing the amplitude of perturbations at some redshift affects the growth at lower redshifts. As we discuss in \ref{error-r}, the constraint on $\mu$ at $z>3$ depends strongly on the width of the bin, i.e. the value of the arbitrarily chosen high-$z$ cutoff. It is also highly correlated with $\mu$ in lower redshift bins and some of the vanilla cosmological parameters.

Unlike $\mu$, the best constrained modes of $\gamma$ do not have a support at high $z$. They peak at low redshifts (bottom two rows in Fig.~\ref{fig:IV_A_mugamma}). This follows from the fact that, according to our definition, to measure $\gamma$ one needs to measure both $\Phi$ and $\Psi$. Therefore, bounds on $\gamma$ come mostly from combining the GC data, which probes $\Psi$ (affected by $\mu$), with the WL signal probing $(\Phi+\Psi)$ (affected by both $\mu$ and $\gamma$). One can also see that there are modes that peak at $z>3$ and at low k (very large scales). This is mainly due to the ISW effect, seen as a contribution to the CMB temperature spectrum, which is sensitive to the time variation of both potentials at all times after the last-scattering.

An important thing to notice is that the best constrained modes of $\mu$ and $\gamma$ show oscillations (nodes) in $k$, but no nodes in the $z$-direction. One has to look at the higher number modes to start seeing oscillations in $z$. For instance, for $\mu$, the first $z$-node appears in the $26$th eigenmode. The number of nodes is indicative of the sensitivity of the function to changes in $k$ and $z$, and we see that the experiments are significantly more sensitive to scale-dependent features of the MG functions, and not as sensitive to the time-dependence. As already pointed out in~\cite{Zhao:2009fn}, this is expected, since the impact of a scale-dependent change in $\mu$ (or $\gamma$) is directly translated into a scale-dependent feature in the WL and GC spectra. For instance, in the case of GC, $\mu$ effectively appears as a scale-dependent normalization factor. On the other hand, the projection of time-dependent features of the MG functions onto the observables involves integration over time which makes detecting $z$-dependent features harder. Also, the amount of information coming from the radial ($z$-direction) is limited by the fact that LSST only probes structures at $z \lesssim 3$, and by the fact that we consider only linear scales, effectively cutting off a significant volume at low $z$.

Fig.~\ref{fig:IV_A_combined} shows the best {\it combined} eigenmodes of $(\mu,\gamma)$. Every combined eigenmode is represented by a pair of values at each point on the $(k,z)$ grid, one value resulting from a variation of $\mu$ on that grid point and the other from a variation of $\gamma$ on the same grid point. We show each eigenmode as a pair of surfaces, one corresponding to $\mu$ and the other to $\gamma$. It can be noticed that the best constrained modes peak at high redshift, which is the result of high sensitivity of the growth to changes in $\mu$ in the high-$z$ bins. Since $\gamma$ does not directly affect the growth rate, the $\gamma$ surfaces of these eigenmodes have a very low amplitude, requiring us to amplify them by a factor of 100 in order to make them visible in the plots. The combined eigenmodes that peak at low redshifts, starting from the 13th, do not exhibit separation of scale or time dependent oscillations, but rather have a diagonal form in the $(z,k)$ space, showing a degeneracy between scale and time. This is because the WL observables dominate the information for combined modes at low redshift.  Indeed,  the changes in the weak lensing kernel due to a shift of the lens along the line of sight ({\it i.e.} a change in redshift) are degenerate with those due to a resizing of the lens ({\it i.e.} a change in scale).

Fig.~\ref{fig:IV_A_Sigma} shows the best eigenmodes for the function $\Sigma$. As it is clear from its definition~(\ref{Sigma}), $\Sigma$ is directly sensitive to the lensing potential $(\Phi +\Psi)$; therefore its signal comes mainly from low redshifts and it does not have well constrained high redshift modes. For this reason the plots for $\Sigma$ span only over the low redshift interval. From these plots we notice a  $k-z$ degeneracy analogous to the one found in the combined $(\mu,\gamma)$ modes as it is clear from a comparison of Fig.~\ref{fig:IV_A_combined} and Fig.~\ref{fig:IV_A_Sigma}.

Finally, Fig.~\ref{fig:IV_A_errors} shows the uncertainties (square roots of covariance eigenvalues) associated with the eigenmodes of $\mu$, $\gamma$, $\Sigma$ and the combined modes. As expected, the combined eigenmodes are constrained best, since they contain all of the information about any departure of either $\mu$ or $\gamma$ from unity. The constraints on $\gamma$ are the weakest, since $\gamma$ is not directly constrained by WL nor GC, while $\mu$ does better than $\gamma$ since it is directly constrained by GC. The eigenmodes of $\mu$ that peak at $z>3$ are constrained better than those of $\Sigma$, while $\Sigma$ is measured better than $\mu$ for modes with support at lower redshifts. This is because at low redshifts most of the information comes from WL, which is directly sensitive to $\Sigma$. As mentioned earlier, the bounds on the $z>3$ bin of $\mu$ are dependent on the arbitrary cutoff and in the next subsection we will marginalize over it.

\subsection{PCA for LSST after marginalizing over high z and galaxy bias, everything else is fixed}
\label{MObias}

\begin{figure}[tbph]
\includegraphics[width=1.05\columnwidth]{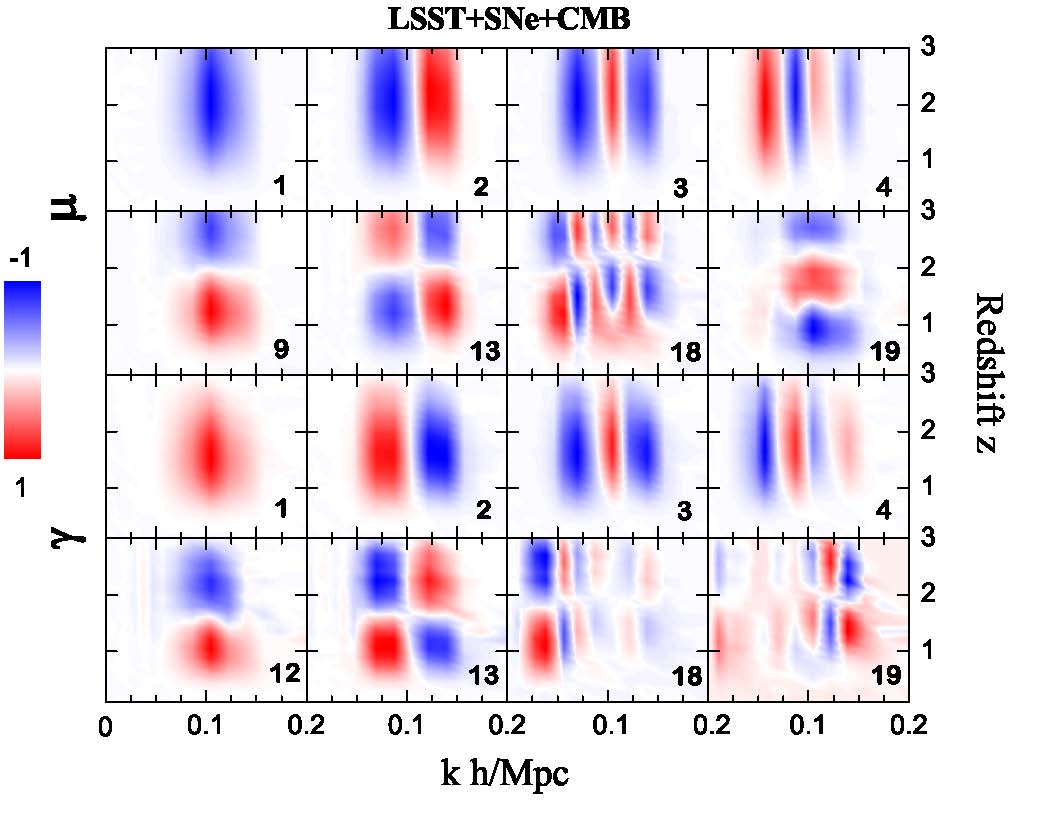}
\caption{The eigenmodes of $\mu$ and $\gamma$ for LSST(+SN+CMB) after marginalizing over the high-$z$ bins, with all other parameters fixed to fiducial values.}
\label{fig:IV_2_B_mugamma}
\end{figure}

\begin{figure}[tbph]
\includegraphics[width=1.05\columnwidth]{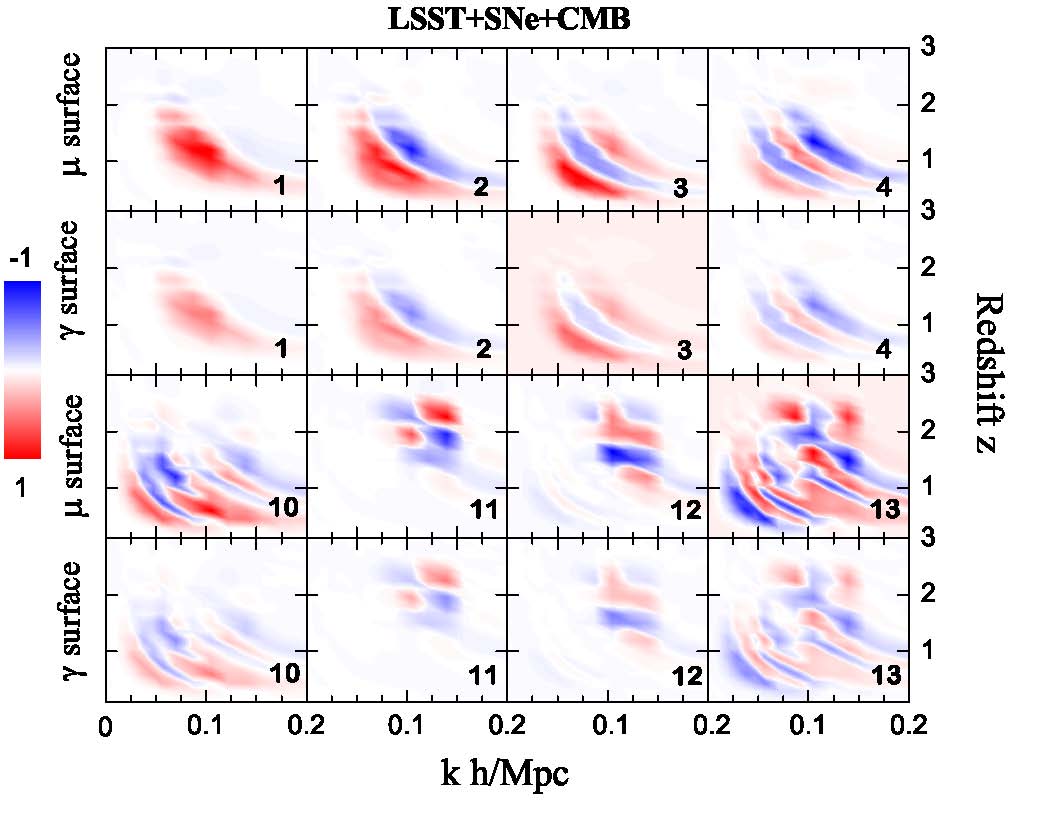}
\caption{The combined eigenmodes of $(\mu,\gamma$) for LSST(+SN+CMB) after marginalizing over the high-z bins, with all other parameters fixed to fiducial values.}
\label{fig:IV_2_B_combined}
\end{figure}

\begin{figure}[tbph]
\includegraphics[width=1.05\columnwidth]{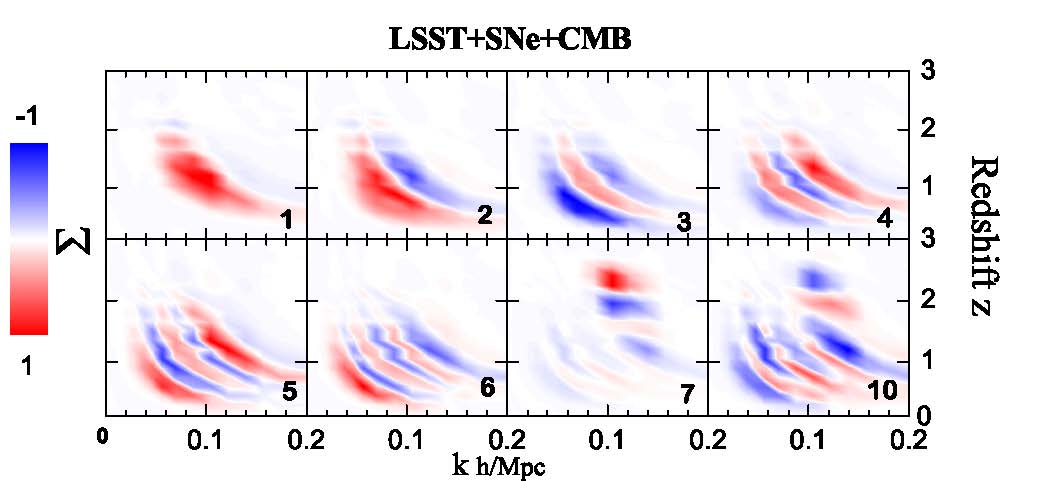}
\caption{The eigenmodes of $\Sigma$ for LSST(+SN+CMB) after marginalizing over the high-z bins, with all other parameters fixed to fiducial values.}
\label{fig:IV_2_B_Sigma}
\end{figure}

In order to remove the dependence of our results on the arbitrary choice of the upper cutoff of the high-$z$ bins ($3<z<30$), we marginalize over the high-z pixels. This is achieved by removing the rows and columns corresponding to high-$z$ pixels from the covariance matrix of the previous subsection, and then diagonalizing it to find the eigenmodes and eigenvalues. This essentially removes all the information about the growth at $z>3$ and, with that, all the $z>3$ features in the eigenmodes. This is seen in Figs.~\ref{fig:IV_2_B_mugamma}-\ref{fig:IV_2_B_Sigma}, where the eigenmodes for $\mu$, $\gamma$, the combined $(\mu,\gamma)$ and $\Sigma$ are shown for this case.

From Fig.~\ref{fig:IV_2_B_mugamma} we notice that, in the absence of the high-$z$ information, the first few best constrained eigenmodes of $\mu$ and $\gamma$ have similar shapes. However, this similarity fades for the higher order eigenmodes, reflecting the different dependences of the two functions on the metric potentials. For instance, the first node in $z$ appears at the $9$th mode for $\mu$ and only at the $12$th for $\gamma$, reflecting a higher sensitivity of $\mu$ to time dependent features.

The best constrained combined $(\mu,\gamma)$ eigenmodes have the same shapes in the $\mu$- and $\gamma$-surfaces, but the $\gamma$-surfaces have a lower amplitude (Fig.~\ref{fig:IV_2_B_combined}). This is again explained by the fact that both functions are constrained by the same experiments (WL and GC), therefore having similar eigenmodes, with $\mu$ being more directly related to the growth of structure.

As can be observed from Fig.~\ref{fig:IV_2_B_Sigma}, after marginalizing over the high-z bins, the best eigenmodes of $\Sigma$ are very similar to the combined $(\mu,\gamma)$ modes. Indeed, once the high-z information is removed, the combined modes are primarily constrained by WL and, therefore, carry more or less the same information as the $\Sigma$ modes.

\begin{figure}[tbph]
\includegraphics[width=1.05\columnwidth]{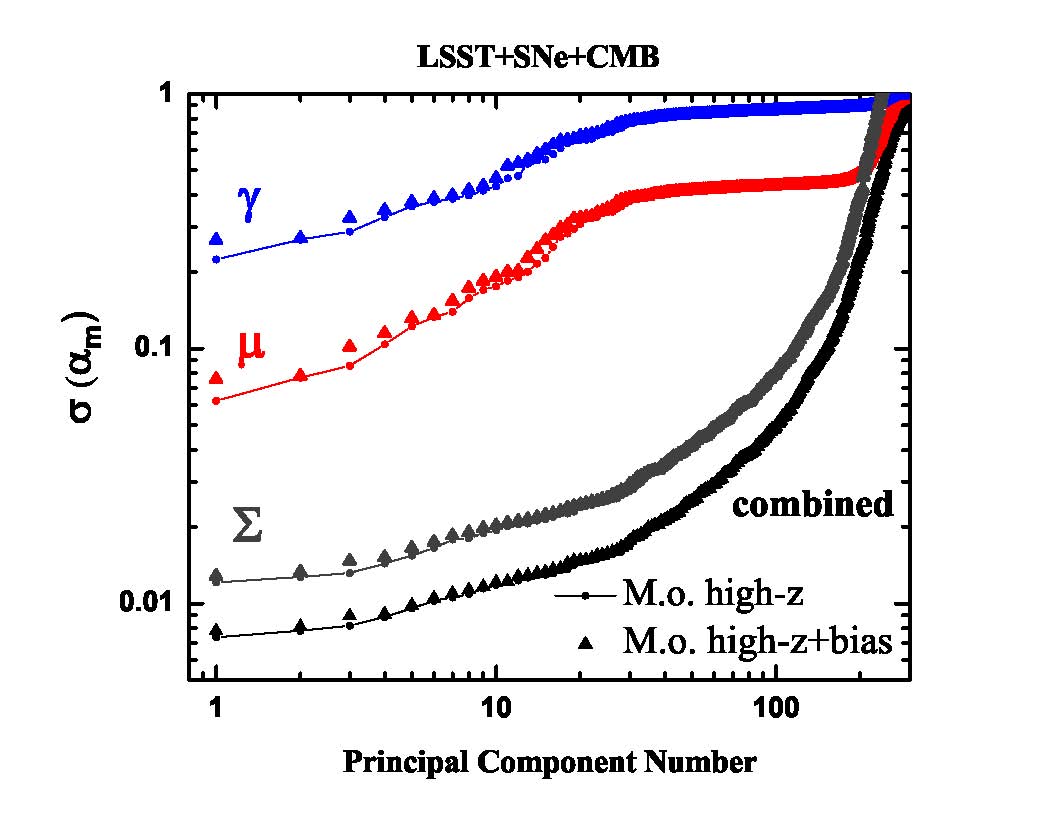}
\caption{The uncertainties associated with the eigenmodes of $\mu$, $\gamma$, $\Sigma$ and the combined $(\mu,\gamma)$ modes for LSST(+SN+CMB), after marginalizing over the high-z bins only (M.o. high-z), and after marginalization over the high-z bins and the galaxy bias parameters (M.o. high-z+bias), with all other parameters fixed to fiducial values.}
\label{fig:IV_2_B_errors}
\end{figure}

\begin{figure}[tbph]
\includegraphics[width=1.05\columnwidth]{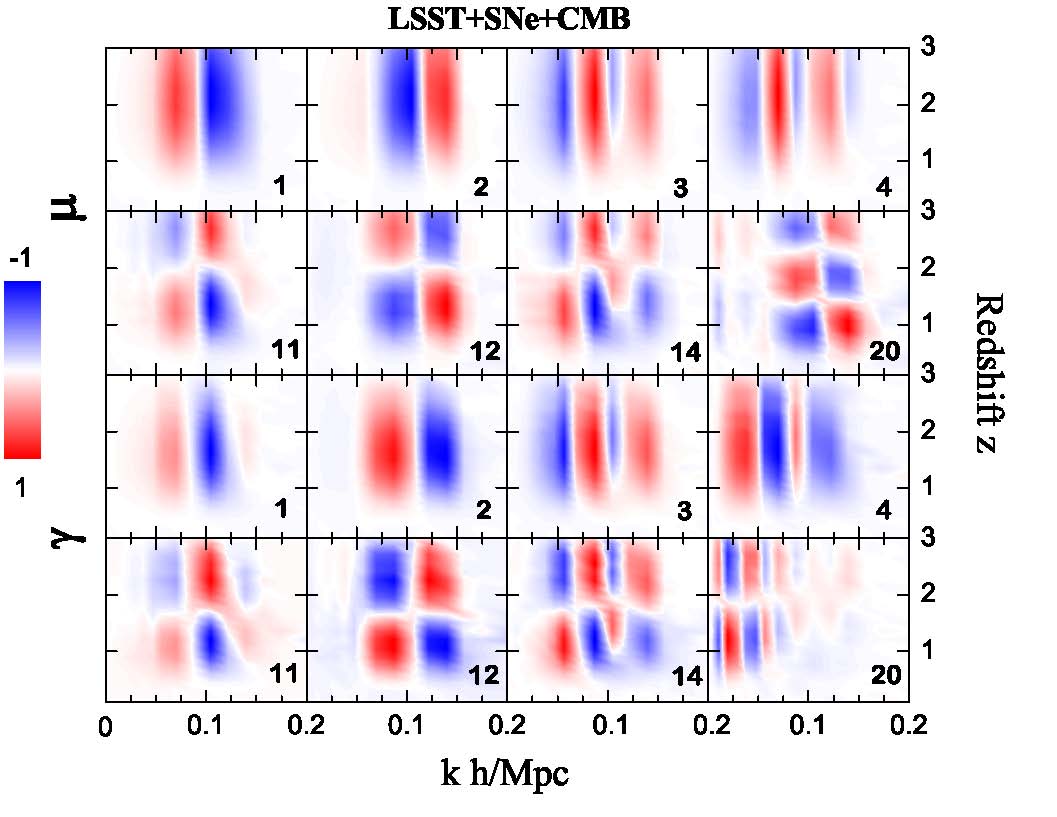}
\caption{The eigenmodes of $\mu$ and $\gamma$ for LSST(+SN+CMB), after marginalizing over the high-z bins and the galaxy bias parameters, with all other parameters fixed to fiducial values.}
\label{fig:IV_2_B3_mugamma}
\end{figure}

Fig.~\ref{fig:IV_2_B_errors} shows the uncertainties associated with the eigenmodes of $\mu$, $\gamma$, $\Sigma$ and the combined $(\mu,\gamma$) modes after marginalizing over the high-z bins. Comparing to Fig.~\ref{fig:IV_A_errors}, we see that now there is no crossing over of the errors on $\mu$ and $\Sigma$, {\it i.e.} all the PCs of $\Sigma$ do better than the $\mu$ ones. This is due to the disappearance of the eigenmodes with high-$z$ support. In addition, we notice a small overall degradation of constraints, which is due to throwing away the information common with the low-z bins.

Next, we marginalize over the galaxy bias parameters. This is achieved by inverting the part of the Fisher matrix that includes the MG pixels and the bias parameters. This effectively removes the information about the overall normalization of $\mu$. This is manifested in the disappearance of some of the eigenmodes, including the homogeneous eigenmodes, e.g. present in Fig.~\ref{fig:IV_2_B_mugamma} but not in Fig.~\ref{fig:IV_2_B3_mugamma}. The same happens to the eigenmodes of $\Sigma$ and of the combination $(\mu,\gamma$) and we do not plot them.

\subsection{PCA for LSST after marginalizing over the vanilla cosmological parameters and $w(z)$.}
\label{MOeverything}
\begin{figure}[tbph]

\includegraphics[width=1.05\columnwidth]{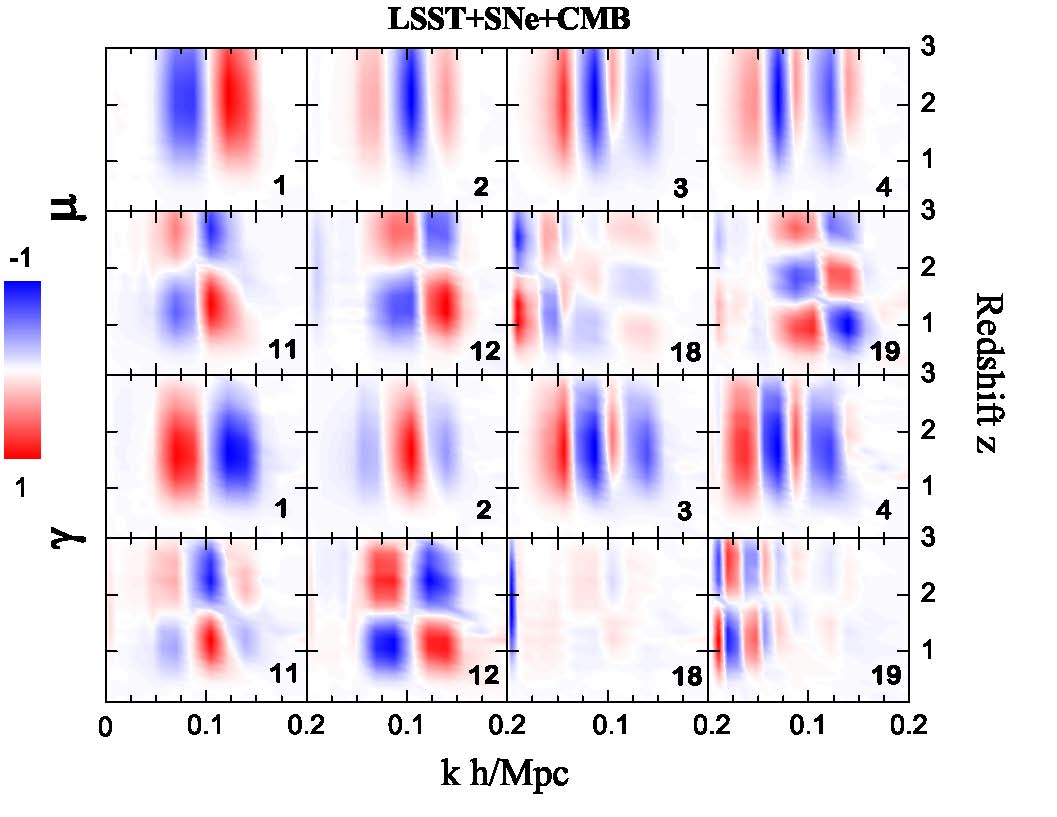}
\caption{The eigenmodes of $\mu$ and $\gamma$ for LSST(+SN+CMB) after marginalizing over all other parameters.}
\label{fig:LSST-mu-gamma-wo_highz-no_sys}
\end{figure}

\begin{figure}[tbph]
\includegraphics[width=1.05\columnwidth]{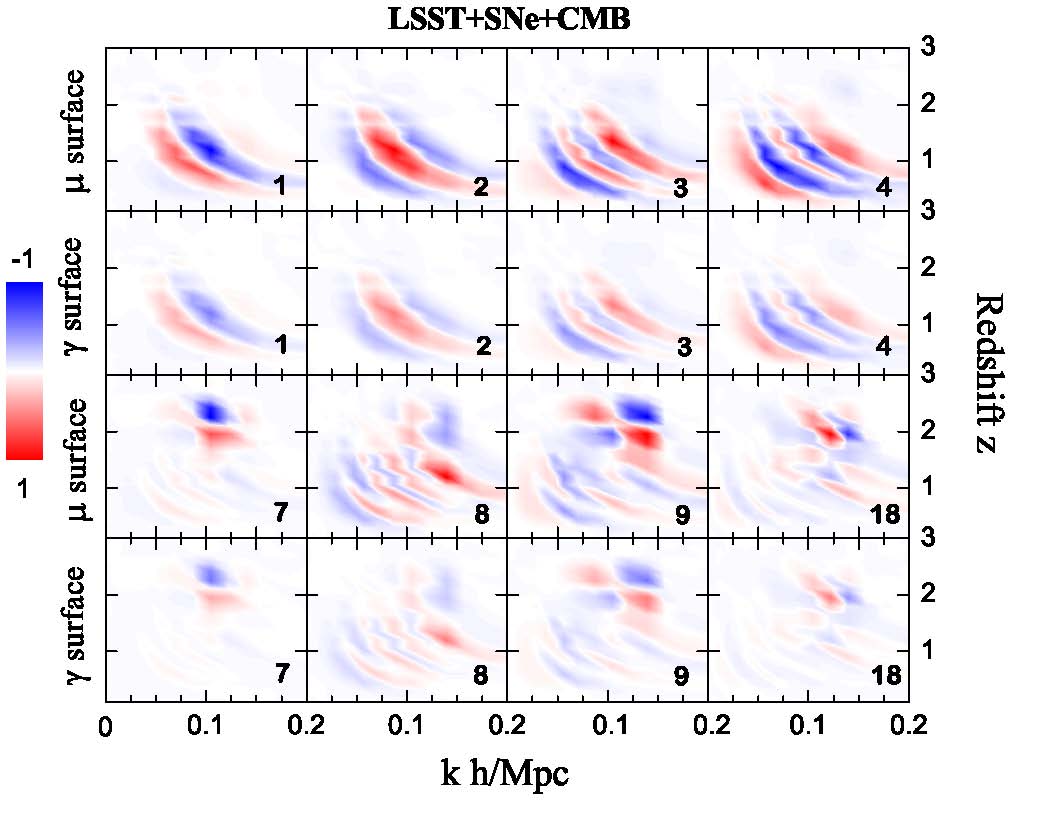}
\caption{The combined eigenmodes of $(\mu,\gamma)$ for LSST(+SN+CMB) after marginalizing over all other parameters.}
\label{fig:LSST-com-wo_highz-no_sys}
\end{figure}

\begin{figure}[tbph]
\includegraphics[width=1.05\columnwidth]{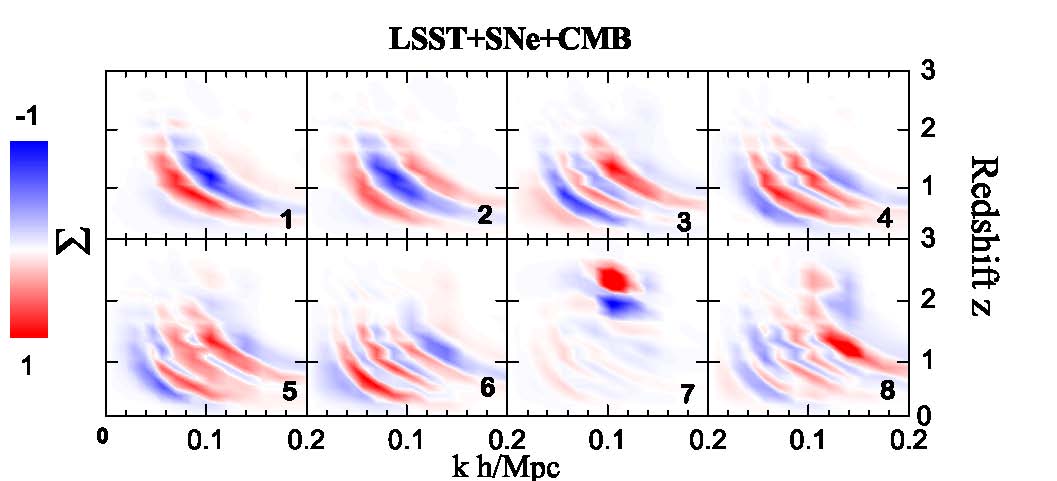}
\caption{The eigenmodes of $\Sigma$ for LSST+SN+CMB after marginalizing over all other parameters.}
\label{fig:LSST-Sigma-wo_highz-no_sys}
\end{figure}

\begin{figure}[tbph]
\includegraphics[width=1.05\columnwidth]{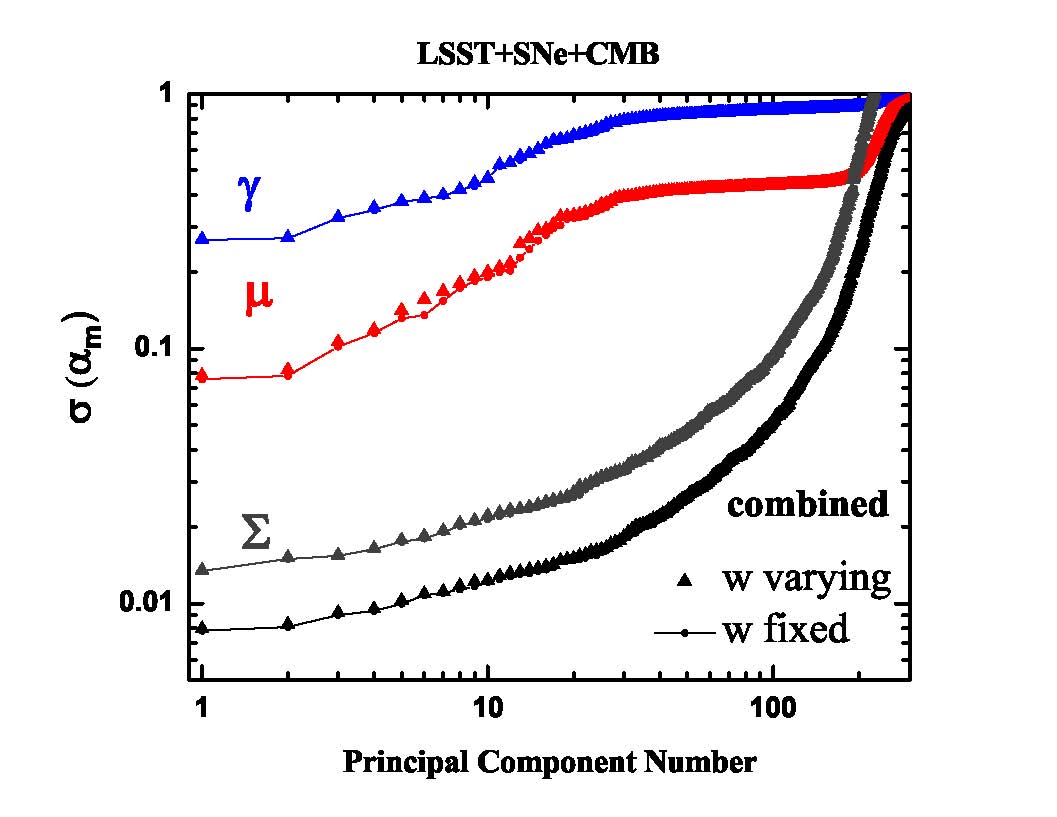}
\caption{The uncertainties associated with the eigenmodes of $\mu$, $\gamma$, $\Sigma$ and the combined $(\mu,\gamma)$ modes for LSST+SN+CMB after marginalizing over all other parameters. Two cases are shown: with $w$ bins fixed to their fiducial value of $-1$ (dots joined by lines), and with $w$ bins varied and marginalized over (triangles).}
\label{fig:LSST-errors}
\end{figure}

We now marginalize over the cosmological parameters:  $\{ \Omega_ch^2, \Omega_bh^2, h, \tau, n_s, A_s\}$ and the binned values of $w(z)$. By doing so we account for the covariance of MG pixels with the vanilla $\Lambda$CDM parameters and the effective dark energy equation of state. Figs.~\ref{fig:LSST-mu-gamma-wo_highz-no_sys}-\ref{fig:LSST-errors} show the PCA results for this case.

The impact of marginalizing over $w(z)$, as opposed to setting it to $w=-1$, is not dramatic and we do not separately show the eigenmodes for the latter case. The associated eignevalues, plotted in Fig.\ref{fig:LSST-errors}, show only a minor differences. This is, in part, due to the high quality of the assumed SNe dataset (see Sec.~\ref{sec:cmbsn}). However, it is also because of marginalizing over the vanilla parameters, the galaxy bias parameters and the high-z bins, which already throws away most of the information that is common between the MG pixels and $w(z)$. The effect of marginalizing over bias parameters and $w$ bins is minor changes in the shape of modes (e.g. second mode in Fig.~\ref{fig:LSST-mu-gamma-wo_highz-no_sys} to be compared with second mode in Fig.~\ref{fig:IV_2_B3_mugamma}), disappearance of some of the modes and an overall degradation of constraints.

\subsection{Constraints on the growth at high redshift}
\label{error-r}

The growth at $z > 3$ is not directly probed by the large scale structure surveys we are considering in this paper. Any modification to growth at $z>3$, such as due to variations of high-$z$ MG pixels, is observed as an overall shift of the amplitude of the growth at all subsequent (i.e. lower) redshifts. This can, in principle, be compensated by counter-variations of the low-$z$ MG pixels. However, we find that bounds on high-$z$ pixels are still relatively tight (at a percent level), even after marginalizing over the low-$z$ bins. Such tight constraints are due to the large width of high-z pixels. Namely, a small change in value of $\mu$ in the high-$z$ bins results in an accumulated modification of growth that can only be compensated by a very large variation of the low-z pixels. On the other hand, the low-$z$ pixels are directly constrained by the surveys and large variations are not allowed. Of course, one can always make the bounds on the high-$z$ bins arbitrarily weak by making the bins narrower.

Since the width of high-$z$ pixels is a somewhat arbitrary parameter, one can ask if another quantity can be introduced to quantify the growth at $z=3$. We take this quantity to be the ratio $r(k)$, defined as:
\begin{equation}
\label{ratio}
r(k) \equiv  \frac{\Delta(z=3,k)}{\Delta (z_{\rm rec},k)} \, ,
\end{equation}
where $\Delta$ is defined in Eq.~(\ref{parametrization-Poisson}) and $z_{\rm rec}$ denotes the redshift at recombination. We can estimate the variance in $r$ ($C_{rr} = \sqrt{\sigma_r^2}$) from
\begin{equation}
\sigma_r = \sum_i \frac{\partial r}{\partial {p_i}} \sigma_{p_i} \, ,
\end{equation}
where $p_i$'s are all the parameters of our model that affect $r$. Our calculation shows that $\sigma_r/r > 1$ for all $k$. In other words, $r$ is completely unbounded, as expected. Note that the calculation would need to be adjusted for DES, for which the highest redshift is $z =1.3$, instead of $z=3$.

\subsection{Comparison with DES}
\label{des-compare}

To get a further insight into how the MG constraints depend on the properties of the experiments, we compare PCA  for LSST+CMB+SNe with that for DES+CMB+SN. That is, we interchange LSST with DES for the WL and GC data, while keeping the assumptions for the CMB and SNe data the same. We perform this comparison only for the case in which we marginalize over all parameters, {\it i.e.} high-z bins, bias parameters, cosmological vanilla parameters and $w(z)$, as in~\ref{MOeverything}.

\begin{figure}[tbph]
\includegraphics[width=1.05\columnwidth]{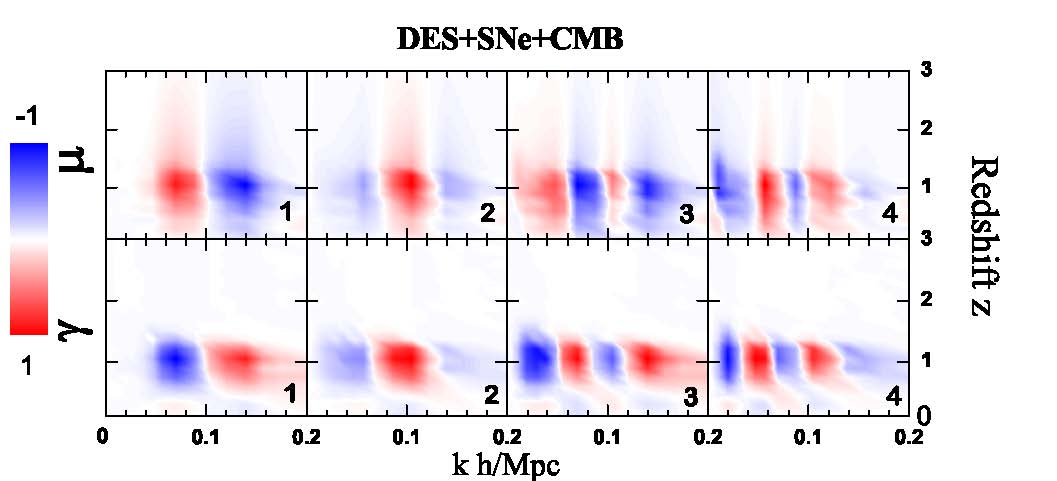}
\caption{The eigenmodes of $\mu$ and $\gamma$ for DES+SN+CMB after marginalizing over all other parameters.}
\label{fig:DES-mu-gamma-wo_highz-no_sys}
\end{figure}

\begin{figure}[tbph]
\includegraphics[width=1.05\columnwidth]{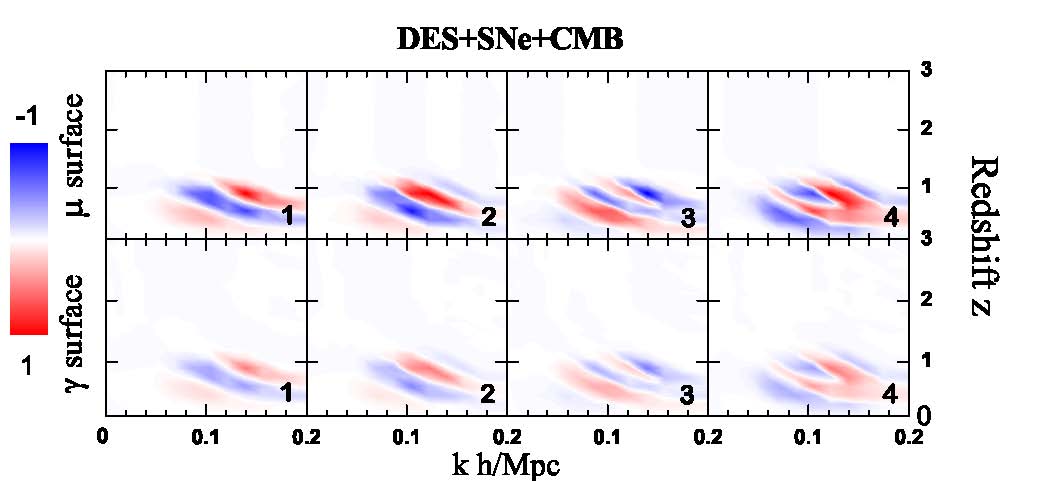}
\caption{The eigenmodes of the combined $(\mu,\gamma)$ modes for DES+SN+CMB after marginalizing over all other parameters.}
\label{fig:DES-com-wo_highz-no_sys}
\end{figure}

Figs.~\ref{fig:DES-mu-gamma-wo_highz-no_sys} and \ref{fig:DES-com-wo_highz-no_sys} show the eigenmodes of $\mu$, $\gamma$ and $(\mu,\gamma)$ for DES+CMB+SN. These results should be compared to Figs.~\ref{fig:LSST-mu-gamma-wo_highz-no_sys} and \ref{fig:LSST-com-wo_highz-no_sys} for LSST. We choose to show a smaller number of eigenmodes, since there are not as many well constrained modes for DES as there are for LSST.

\begin{figure}[tbph]
\includegraphics[width=1.05\columnwidth]{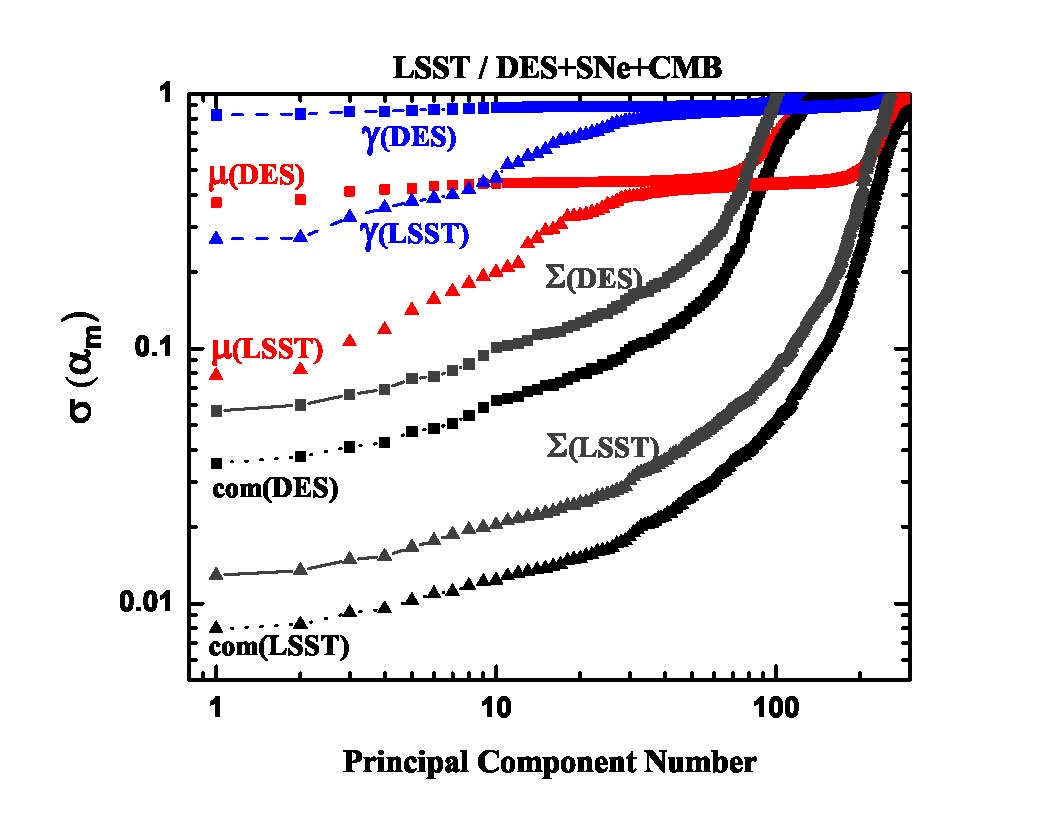}
\caption{The uncertainties associated with the eigemodes of $\mu$, $\gamma$, $\Sigma$ and combined $(\mu,\gamma)$ modes for DES(+SN+CMB) and for LSST(+SN+CMB) errors, after marginalization over all other parameters.}
\label{fig:errors-full}
\end{figure}

As can be seen from Figs.~\ref{fig:DES-mu-gamma-wo_highz-no_sys} and \ref{fig:DES-com-wo_highz-no_sys}, the range over which the eigenmodes vary is limited to smaller redshifts ($z < 1$), which reflects the redshift range probed by DES. Like in the case of LSST, there is a higher sensitivity to scale-dependent features. Furthermore, in the case of DES, there are modes with oscillations in $z$. Some of the combined eigenmodes are also absent for DES . Fig.~\ref{fig:errors-full} shows a comparison of the uncertainties associated with the modes for LSST and DES.

While the overall sensitivity of DES is less than LSST by a factor of few, DES, when combined with CMB and SNe data, is still able to constrain several eigenmodes with better than $10$\% accuracy. We may not be able to detect a time-dependent MG feature with high confidence from DES, but it is possible to constrain scale-dependent features.

\section{Effects of systematic errors}
\label{sec:systematics}

Systematics are notoriously hard to predict, and forecast results that include systematics are in general sensitive to the modeling and the assumed priors~\cite{Huterer:2005ez,Zhan:2008jh,Kirk:2011sw,Laszlo:2011sv}. Here we consider some of the systematics that will affect future tomographic surveys and study their impact on our PCA results. This can give us a general insight on how systematic errors could affect our inferences of the eigenmodes and the corresponding uncertainties. The systematics we consider here are the photo-$z$ errors and some of the errors in the measurement of the point spread function (PSF). These errors are modeled in~\cite{Huterer:2005ez,Zhan:2008jh} and we use their parametrization for our Fisher analysis. 

The effects of the systematics are detailed in the Appendix, and with these assumptions we study the degradation of ability of DES and LSST to constrain MG by marginalizing over the systematics parameters without applying any prior.  As we show later, the degradation is apparent, but not disastrous. It is true that the catastrophic photo-$z$ errors (CPE) for WL surveys~\cite{Bernstein:2009bq} can lead to disastrous degradation on the constraints of the cosmological parameters, but as shown in \cite{Bernstein:2009bq}, an additional 30,000 spectroscopic redshifts can help to control the bias in cosmological parameters due to CPE under the level of statistical errors. The most significant effect is to reduce the ability of LSST to detect the $z$-dependence of $\mu$. This is a preliminary analysis, which does not include scale-dependent systematics, which can be particularly important in MG studies (as was found in~\cite{Zhao:2010dz} where PCA was applied to a set of existing data including CFHTLS). 

\begin{figure}[h]
\includegraphics[width=1.\columnwidth]{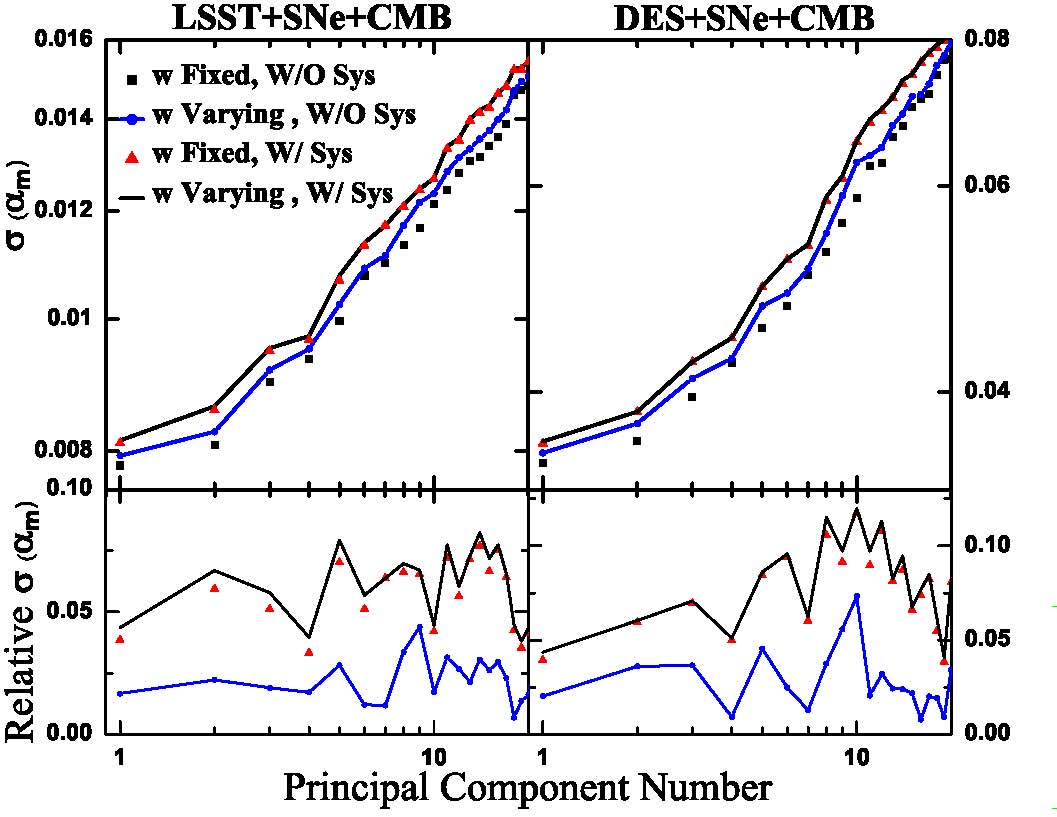}
\caption{Top: The uncertainties associated with the combined $(\mu,\gamma)$ eigenmodes for LSST (left) and DES (right) for four cases: (I) without systematics and $w$ fixed, (II) without systematics but $w$ varied, (III) with systematics but $w$ fixed and (IV) with systematics and $w$ varied. Bottom: The uncertainties for LSST (left) and DES (right) relative to Case 1. }
\label{fig:sys-comparison-com}
\end{figure}

\begin{figure}[h]
\includegraphics[width=1.\columnwidth]{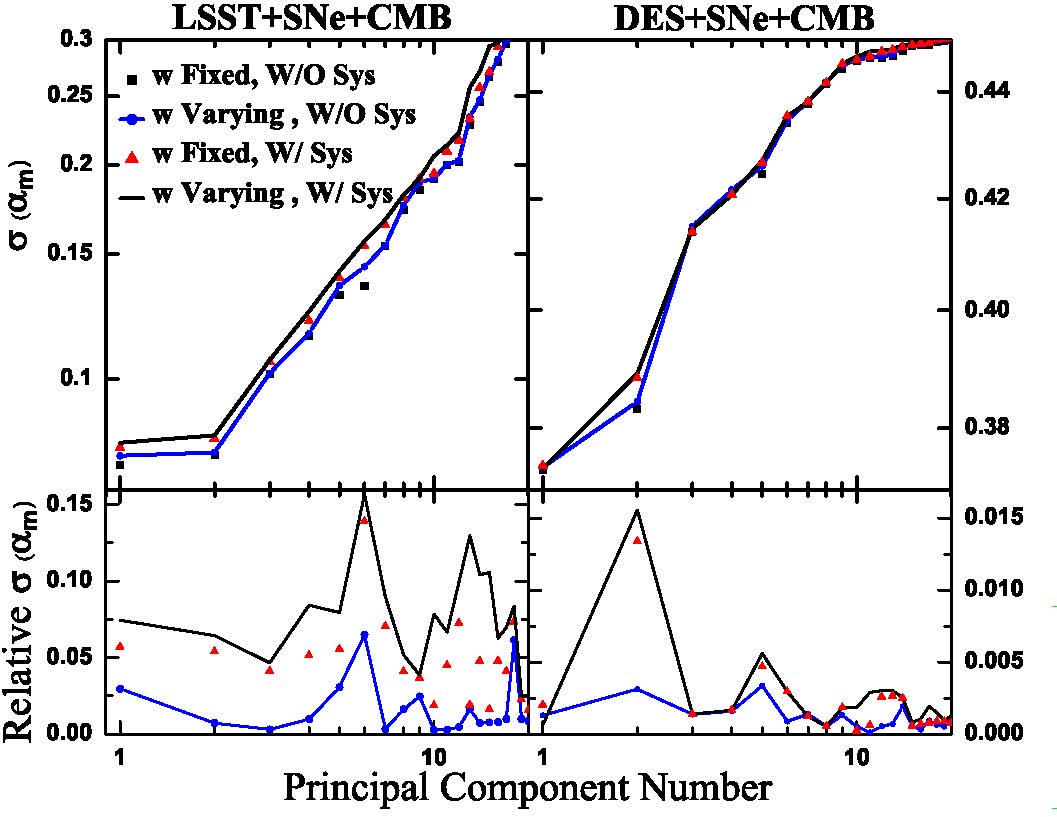}
\caption{Top: The uncertainties associated with eigenmodes of $\mu$ for LSST (left) and DES (right) for four cases: (I) without systematics and $w$ fixed, (II) without systematics but $w$ varied, (III) with systematics but $w$ fixed and (IV) with systematics and $w$ varied. Bottom: The uncertainties for LSST (left) and DES (right) relative to Case 1. }
\label{fig:sys-comparison-mu}
\end{figure}

Let us look at the errors for combined eigenmodes shown in Fig.~\ref{fig:sys-comparison-com} for the following cases: (I) no systematics, with $w$ fixed; (II) no systematics, with $w$ varied; (III) with systematics, and $w$ fixed; (IV) with systematics, and $w$ varied. We can see that allowing for systematics degrades the constraints more than allowing for variations in $w$. Also, once the systematic errors are included, allowing $w$ to vary does not degrade the constraint further. This means that the MG pixels are more degenerate with the systematics parameters than the $w$ bins, implying that the uncertainty in the galaxy distributions, which is basically photo-$z$ error from the systematics, can affect the constraint on the growth more significantly than $w$ does. In general, the errors on the best constrained modes are degraded by $\lesssim 10\%$ for both LSST and DES. Another interesting observation is that the largest degradation does not happen for the first few modes, but for the intermediate modes. This is reasonable -- the first few modes do not have nodes in $z$ and thus are relatively immune to the systematics dominated by the photo-$z$ errors.

We have shown the errors on the eigenmodes for $\mu$ in Fig.~\ref{fig:sys-comparison-mu} where the same cases as above are considered. One notices that degradation on $\mu$ is less than what we had for combined case, especially no significant changes for DES are found. For $\gamma$, degradation is very small simply because the constraints on $\gamma$ eigenmodes are very weak in the first place. We are not therefore showing $\gamma$ errors here.

\begin{figure}[h]
\includegraphics[width=1.05\columnwidth]{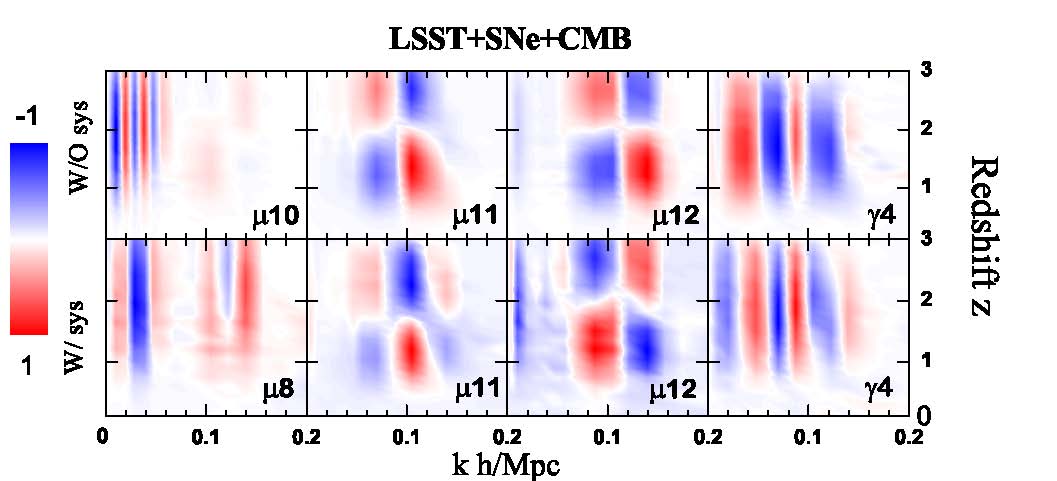}
\caption{The eigenmodes for LSST(+SN+CMB). The upper(lower) panel corresponds to the case without(with) systematics. The $10$th, $11$th and $12$th modes of $\mu$ without systematics, are compared to the $8$th, $11$th and $12$th modes of $\mu$ with systematics respectively. These modes are chosen since they correspond to the last z-independent and first z-dependent mode respectively, in the two case (with and without systematics). Analogously, the $4$th eigenmode of $\gamma$ without systematics is compared to the $4$th mode of $\gamma$ with systematics. It is an illustration of how LSST eigenmodes are distorted as a result of accounting for systematics.}
\label{fig:LSST-sys-comparison}
\end{figure}

Another observation is that systematics can create new, or destroy existing modes, so that the modes with the same order in the PCA sequence in the cases with and without systematics can be different modes.   One can see this clearly by looking at the eigensurfaces in Fig.~\ref{fig:LSST-sys-comparison}, where we illustrate the three $\mu$ modes and one $\gamma$ mode with and without systematics. They look similar except that the modes with systematics in general have more nodes in $k$, indicating that systematics do not just dilute
the constraints on the old modes, but also make some modes disappear (or make them very poorly constrained). The general trend is that eigenmodes with very high frequency features in $k$ are no longer well-constrained after inclusion of systematics.

\begin{figure}[h]
\includegraphics[width=1.05\columnwidth]{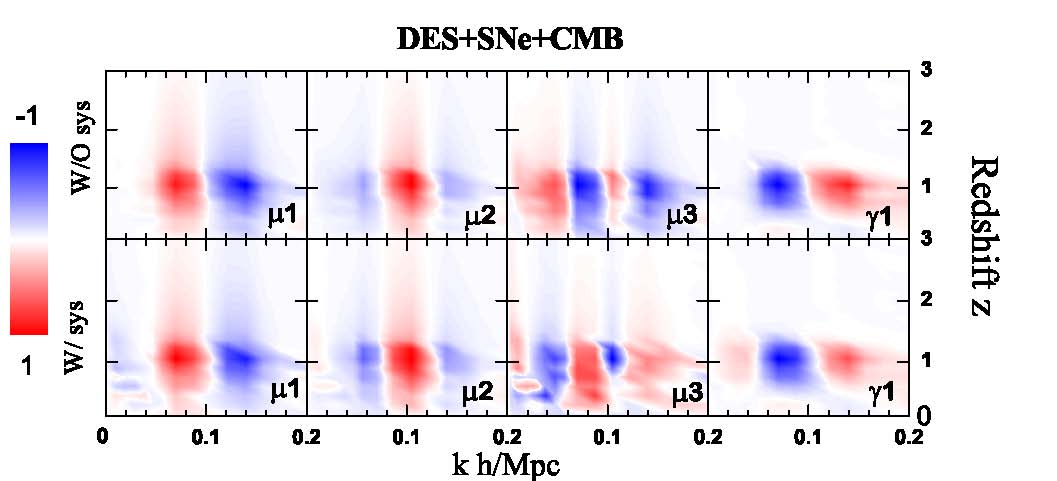}
\caption{The eigenmodes for DES(+SN+CMB). The upper(lower) panel corresponds to the case without(with) systematics. The three modes of $\mu$ and first mode of $\gamma$ without systematics, are compared to corresponding modes of $\mu$ with systematics. It is an illustration of how DES eigenmodes would be distorted due to systematics.}
\label{fig:DES-sys-comparison}
\end{figure}

Overall, we find that the inclusion of systematics results in a noticeable, but not dramatic, dilution of the constraints on MG from DES. This is because photo-$z$ errors would most immediately affect the $z$-dependence of MG, to which DES was already mostly insensitive even without systematics. The main constraints from DES, as can be seen from the shape of the eigenmodes, will be on the scale-dependence of $\mu$ and $\gamma$, and that information is somewhat reduced, but mostly preserved. The impact of the systematics on LSST could be more significant, simply because LSST has a higher potential for resolving $z$-dependent features. In this preliminary analysis, we find that allowing for systematic errors under the assumptions of \cite{Huterer:2005ez,Zhan:2008jh} preserves most of the scale-dependent information from LSST, but decreases our ability to measure eigenmodes of $\mu$ with $z$-dependent features.

\section{Degeneracy between MG parameters and dark energy EoS}
\label{sec:w}

\begin{figure}[tbp]
\includegraphics[width=1.\columnwidth]{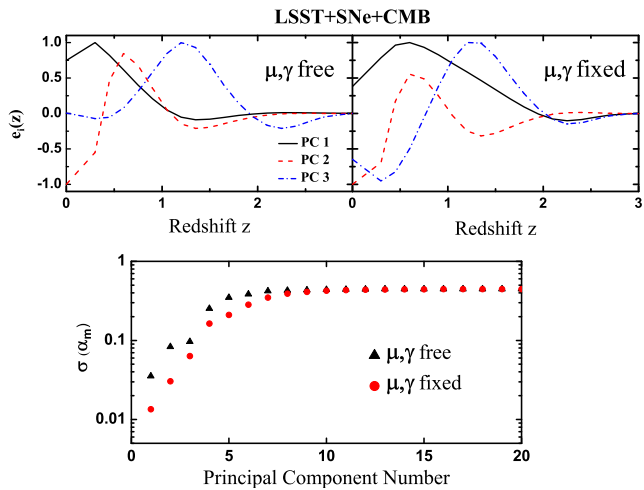}
\caption{Top: First three eigenmodes of $w(z)$ with (left) and without (right) MG included. Bottom: Errors for $w(z)$ eigenmodes with and without MG included (there is a prior of $0.5$ on $w$ bins). } \label{fig:w-PCA}
\end{figure}

In Sec.~\ref{MOeverything}, we marginalized over the binned equation of state and analyzed the impact of this marginalization on the eigenmodes of $\mu$ and $\gamma$. Here we investigate how marginalizing over MG pixels affects the eigenmodes and eigenvalues of $w(z)$. In this case, we diagonalize the block of the covariance matrix containing the 21 $w$-bins. Fig.~\ref{fig:w-PCA} (top) shows the best three eigenmodes of $w$ in the case when $\mu$ and $\gamma$ pixels are co-varied, and when they are fixed to their fiducial values.

The most noticeable effect of the marginalization is a reduction of the amplitude of the best constrained modes at higher z. In other words, letting MG parameters vary squeezes the best constrained eigenmodes of $w(z)$ towards low redshift. This is expected, since most of the information on $w(z)$ comes from the SNe and from the probes of LSS at higher redshift. The latter is largely degenerate with MG, and so the high-$z$ information on $w(z)$ is erased. On the other hand, LSS does not contribute much to the low-z information, since we restrict to modes in the linear regime, therefore the marginalization has little impact on the low-redshift parts of the $w(z)$ modes.

Overall the effects of the marginalization over the MG functions are not dramatic and future surveys will have the ability to measure {\emph{both}} $w$ and MG.  Fig.~\ref{fig:w-PCA} (bottom) shows the degradation of errors on the $w$ eigenmodes after marginalizing over other parameters, including MG. Note that a prior of $\sigma_i = 0.5$ has been put on the bins of $w$.

\section{Projected errors on parameters of specific models from principal components}
\label{sec:projection}

MG pixels can be viewed as a compressed form for the information we get from observation about the linear growth. One can perform a PCA and store this information in the eigenmodes and their eigenvalues. The advantage of PCA is that the information can be compressed and used more efficiently, namely, the well-constrained eigenmodes usually carry almost all of the existing information.

Exploiting information stored in the eigenmodes, we can easily emulate any other parameterization to forecast parameter errors without regenerating the Fisher matrices from scratch~\cite{Crittenden:2005wj}. In other words, we can treat MG pixels as our observables and use them to calculate the Fisher matrices for specific model parameters.

The Fisher matrix can be written as
\begin{equation}
F_{ab} = \sum_{\alpha \beta}  {\partial O_{\alpha} \over \partial p^a} C^{-1}_{\alpha \beta}  {\partial O_{\beta} \over \partial p^b} \ ,
\end{equation}
where $O_{\alpha}$ are cosmological observables and $p^a$ are parameters of a specific model. This can be rewritten as
\begin{eqnarray}
\label{fisherinfisher}
F_{ab} &=& \sum_{ij}  {\partial \mathcal{P}_i \over \partial p^a} \left[ \sum_{\alpha \beta}   {\partial O_{\alpha} \over \partial \mathcal{P}_i} C^{-1}_{\alpha \beta} {\partial O_{\beta} \over \partial \mathcal{P}_j}\right]  {\partial \mathcal{P}_j \over \partial p^b} \ , \nn \\
&=& \sum_{ij} {\partial \mathcal{P}_i \over \partial p^a}~ F_{ij}~ {\partial \mathcal{P}_j \over \partial p^b}
\end{eqnarray}
where $\mathcal{P}$'s are MG pixels and $F_{ij}$ is the $ij$ element of their Fisher matrix.
All we need now is to expand the derivatives of the MG pixels with respect to a given parameter in the eigenmode basis. That is, for each of the new parameters, we find the coefficients $\mathcal{C}^{a}_{l}$ such that \footnote{Note that since we are working on a discrete ($k,z$) grid, our equations are in a discrete form. Analogous expressions could be written for the continuous case. For example, Eq.~(\ref{partials}) for a continuous $\mu$ function would be ${\partial {\mu(a,k)} \over \partial p^a} = \sum_l \mathcal{C}^{a}_{\mu,l}~ {e}^{\mu}_l(a,k)$.}
\begin{equation}
\label{partials}
{\partial{\mathcal{P}_i} \over {\partial p^a}} = \sum_l \mathcal{C}^{a}_{l} ~ {e}^{l}_i \ ,
\end{equation}
where the sum is over all the eigenmodes and ${e}^{l}_i$ is value of the $l$th eigenmode at the $i$th pixel of the 2D ($k,z$) grid. Substituting Eq.~(\ref{partials}) in Eq.~(\ref{fisherinfisher}), we can then find the Fisher matrix for the parameters of our model by simple projection as
\begin{equation}
\label{fisherprojected}
F_{ab} = \sum_i \mathcal{C}^a_i ~ \mathcal{C}^b_i ~ \lambda_i^{-1} \ ,
\end{equation}
where again the sum is over all the eigenmodes, and $\lambda_i$'s are the corresponding eigenvalues of the covariance matrix for MG pixels.

We now illustrate the details of this method by applying it to a one-parameter model which gives a good approximation of $f(R)$ theories in the quasi-static limit~\cite{Giannantonio:2009gi} (and which is a customized form of the more general parametrization introduced in~\cite{BZ08})
\begin{eqnarray}
\label{B0}
\mu(a,k)&=&\frac{1}{1-1.4 \cdot 10^{-8} \lambda^2a^3}\frac{1+\frac{4}{3}\lambda^2\,k^2a^4}{1+\lambda^2\,k^2a^4}\,, \nn \\
\gamma(a,k)&=&\frac{1+\frac{2}{3}\lambda^2\,k^2a^4}{1+\frac{4}{3}\lambda^2\,k^2a^4}\,,
\end{eqnarray}
where $\lambda^2=B_0\,c^2/(2H_0^2)$ and $c$ is the speed of light. The parameter $\lambda$ is the mass of the $f(R)$ scalar degree of freedom today. In \cite{Schmidt:2009am,Lombriser:2010mp}, a bound of $B_0 \lesssim 10^{-3}$ at $95$\% confidence level was found based on the current cluster abundance data, which extends to mildly non-linear scales.

Here, we forecast the constraints on $B_0$ from LSST, based only on linear scales, in two ways: using a direct Fisher matrix calculation and a Fisher matrix projection described above. We choose $B_0 = 0$, corresponding to GR, as the fiducial model. We use the same combination of future data for the direct Fisher matrix calculations as we did for the PCA of MG, and use the eigenmodes and eigenvalues of Section~\ref{MOeverything} for the Fisher matrix projection. The derivatives on the left hand side of  Eq.~(\ref{partials}) can be calculated analytically from Eq.~(\ref{B0}). Since we are working on a 2D grid of $k$ and $a$, we use the averaged value of these derivative expressions over each pixel and compute the expansion coefficients of eq.~(\ref{partials}) numerically. Table \ref{tab:B0} shows the forecasted constraints on $B_0$ from LSST and DES combination data obtained using the two methods. The results show a reasonable agreement.

\begin{table}[tbph]
\begin{tabular}{|cc|cc|}
\hline
\multicolumn{2}{|c|}{DES} & \multicolumn{2}{c|}{LSST}\\
\hline
Direct & Projection  &  Direct & Projection   \\
\hline
$1.5 \cdot 10^{-6}$  & $2.5 \cdot 10^{-6}$ & $3.1 \cdot 10^{-7}$ & $ 2.4 \cdot 10^{-7}$  \\
\hline
\end{tabular}
\caption{Error forecasts for the $B_0$ parameter for DES and LSST in combination with CMB and SNe data.The results for Fisher matrix projection formalism used here is compared to the direct Fisher matrix calculation. }
\label{tab:B0}
\end{table}

One should be careful when working with projection method. For example, the priors used for calculating the covariance matrix for MG pixels ($C_{ij}$ in eq.~(\ref{fisherinfisher})) should be the same as the priors that would be used for a direct fisher analysis for a model. Ideally, the fiducial models should also be the same in both approaches.

As mentioned before, one of the advantages of the PCA approach is that we are able to compress information using only the best principal components. We find that in order to reproduce the errors shown in the ``Projection'' columns of Table~\ref{tab:B0} at about $95$\% precision we only need $\sim 25$\% of the eigenmodes. Such compression of information can be useful, given the increasing volume of cosmological data.

\section{Summary and Outlook}

In this paper we have extended the principal component analysis first performed in~\cite{Zhao:2009fn}.
As shown in previous sections, upcoming and future weak lensing surveys will provide high precision data on the relationships between matter overdensities, curvature of space and the Newtonian potential, offering an unprecedented opportunity to test GR on cosmological scales. 

In Section~\ref{sec:formalism} we have introduced the MG functions needed to parametrize the evolution of cosmological perturbations on linear scales. As discussed in~\cite{Pogosian:2010tj}, there is not a unique choice of these functions, and we have presented results for two alternative choices: the pair used in~\cite{Zhao:2009fn}, with $(\mu,\gamma)$ encoding deviations in the Poisson and anisotropy equations, respectively, as well as the pair $(\mu,\Sigma$), with $\Sigma$ in Eq.~(\ref{Sigma}) directly related to the WL potential. The main benefit of using these functions is that they allow for a model-independent test of  the growth dynamics on cosmological scales even though they do not necessarily have a simple form in specific models of MG. Quite generally, in fact, they are defined through {\it solutions} of the equations of motion and depend on the choice of the initial conditions; still one can store observable information in these functions in a model independent way, and simply use care when translating the findings on $\mu$ and $\gamma$ into results on the parameters of specific models~\cite{Zuntz:2011aq}. In Sec.~\ref{sec:formalismB} we have given a detailed description of how to perform the two-dimensional PCA of these functions with the aim of offering a useful technical reference for anyone who wishes to apply PCA to modified growth. In the same spirit, we have reviewed the observables and surveys used for our analysis in Sec.~\ref{sec:observables}. 

The bulk of the results is presented in Sec.~\ref{sec:mgpca}, where we have analyzed the principal components (eigenmodes) of the MG functions for the combination of WL survey, CMB and SNe experiments detailed in Sec.~\ref{sec:experiments}. As already noted in~\cite{Zhao:2009fn}, the number of well-constrained eigenmodes gives a forecast of how many degrees of freedom describing deviations from GR will be constrained, and is particularly informative when comparing the outcome for different surveys. The shapes of the eigensurfaces indicate the regions of scale and redshift where the surveys under consideration will be most sensitive to departures from GR. 

We have given a detailed presentation of the eigenmodes and eigenvalues of the functions $\mu$, $\gamma$, $\Sigma$ as well as of the combination $(\mu,\gamma)$, comparing them and interpreting the differences. We have studied thoroughly the degeneracy between the MG functions and other cosmological parameters by progressively varying and marginalizing over the different parameters. At every step we have interpreted and explained the effects of the marginalizations. Of particular interest is the analysis of the degeneracy with the equation of state. We have found that after marginalizing over the MG functions, the high-z information on $w(z)$ is erased and its best constrained eigenmodes  are squeezed towards low redshift; however, as we show in Sec.\ref{sec:w}, the effects of the marginalization are not dramatic and future surveys will have the ability to constrain \emph{both} $w$ and modified growth.
From the comparison of the results for LSST and DES, in Sec.~\ref{des-compare}, we notice that LSST will have overall a higher sensitivity to modified growth and will be more sensitive to time-dependent features. In Sec.~\ref{sec:systematics}, we studied the effects of WL systematics for LSST, and found that the degradation of constraints on MG is not significant, at least of the systematics models we have considered, and especially after one marginalizes over an arbitrary $w(z)$.

Finally, we have shown the utility of the PCA approach as a data compression stage. One can store the information contained in observables in terms of the MG pixels, or the eigenmodes of two functions. One can later use this information to project on to constraints on the parameters of specific models. For example, in Sec.~\ref{sec:projection}, we projected the errors on the MG functions to forecast the error in $B_0$. We have shown that only a fraction of the total number of eigenmodes  is needed to obtain this constraint.

The degeneracy between $\mu$ and $\gamma$ or $\Sigma$ can be further broken by adding information from redshift space distortions (RSD) measurements~\cite{Song:2010fg}. It will be interesting to extend our analysis to include RSD for surveys that simultaneously measure lensing and peculiar velocities, such as DESpec and Euclid. 

Another direction for future work is to revisit the assumed scale-independence of the galaxy bias. In \cite{Hui:2007zh}, it was shown that a scale-dependent growth necessarily implies a scale-dependent bias on linear scales. In the present analysis, this was effectively encoded in our function $\mu$, but in a future study it may be interesting to include the scale-dependent bias explicitly. In addition to scale-dependent bias, scale-dependent initial conditions are also likely to be degenerate with the MG modes (and probably worse, as they impact lensing as well.)
Arguably, this is a more likely degree of freedom than the MG scale dependence. 

Our technique represents a model-independent way of analyzing the power of cosmological surveys to constrain modified growth. In addition to forecasts, it can be applied to current data. For instance, given an array of experiments, one can use a Fisher forecast to first find the eigenmodes, then fit the amplitudes of these modes to real data. Because these modes are expected to be (nearly) orthogonal to each other, it does not matter if one fits them one by one or simultaneously. If any of them is found to deviate from zero significantly, it would constitute a smoking gun for modified gravity. Alternatively, if Fisher forecasted eigenmodes are found to be non-negligibly correlated, it would indicate that the fiducial model assumed in their derivation was
wrong and that a modification is needed. One may also attempt reconstructing $\mu(k,z)$ and $\gamma(k,z)$ from data, using the correlated prior technique introduced in \cite{Crittenden:2005wj, Crittenden:2011aa} for reconstructions of $w(z)$.

\acknowledgments  

We acknowledge helpful discussions with Tessa Baker, Pedro Ferreira, Fabian Schmidt, Constantinos Skordis. LP thanks the Institute of Cosmology and Gravitation in Portsmouth for hospitality. The work of AH and LP is supported by an NSERC Discovery Grant. GZ, RC and KK are supported by STFC grant ST/H002774/1. AS is supported by NSF grant No. AST-0708501. KK is also supported by the ERC and the Leverhulme trust.

\appendix

\section{The model for WL and GC systematics}
\label{sec:appendix}

We follow \cite{Huterer:2005ez,Zhan:2008jh} and consider three sources of systematics for future tomographic imaging surveys: photo-$z$ errors, as well as additive and multiplicative errors due to the uncertainty of the point spread function (PSF) measurements.

\subsubsection{Photo-$z$ Errors}

The redshift errors may stem from three sources: the distortion of
the total galaxy distribution, $z$-bias and $z$-scatter. Suppose the
distribution of the total galaxy is \be\label{gal_tot}
\bar{N}(\bar{z})\propto\bar{z}^2{\rm exp}(-(z/z_0)^2) \ee where
$\bar{z}$ denotes the true, spectroscopic redshift (throughout we
use the over-bars to denote the precisely measured quantities to any
precision), and $z_0$ is the median redshift for a given survey.
Then the galaxy distribution of the $i$th bin is,
\be\label{gal_bin1}
\bar{N}_i(\bar{z})\propto{\bar{N}(\bar{z})}\Big[{\rm
erfc}\Big(\frac{\bar{z}_{\rm
i-1}-\bar{z}}{\sqrt{2}\bar{\sigma}(\bar{z})}\Big)- {\rm
erfc}\Big(\frac{\bar{z}_{\rm
i}-\bar{z}}{\sqrt{2}\bar{\sigma}(\bar{z})}\Big)\Big] \ee
where $\sigma(z)$ denotes the photo-$z$ error at redshift $z$, and erfc is the complimentary error function.

\begin{figure}[htp]
\hspace*{-0.8cm}\includegraphics[width=1.2\columnwidth]{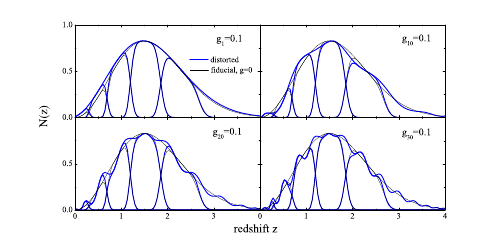}
\caption{Redshift error - distortion: distortion of the total galaxy distribution
due to uncertainties in redshift measurement, plotted here for LSST. $g_i$'s are 
expansion coefficients of $\Delta\bar{z}$ in terms of Chebyshev polynomials (\ref{chy}).} \label{fig:chb}
\end{figure}

{\it \underline{Distortion of the total galaxy distribution:}}
Suppose the observers measure the redshift $z$ using some
photometric method, and $\Delta\bar{z}\equiv{z-\bar{z}}$ denotes the
error. This error, in general, might induce a distortion of the
overall distribution of the galaxies. To quantify this effect, we
follow \cite{Huterer:2005ez} to expand $\Delta\bar{z}$ using the
smooth Chebyshev polynomials ($T_n(x)={\rm cos}(n{\rm
arccos}~x)$): \be\label{chy} \Delta\bar{z}=\sum_{i=1}^{N_{\rm
chb}}g_iT_i\Big(\frac{\bar{z}-z_{\rm max}/2}{z_{\rm max}/2}\Big). \ee
As argued in \cite{Huterer:2005ez}, choosing $N_{\rm chb}=30$ is
large enough to yield convergent result, and we follow this setting.
If the expansion coefficients $g_i$'s are much smaller than unity,
the biased galaxy distribution can be estimated as
\be\label{gal_tot_bias}
N(z)=\bar{N}\Big[z-g_iT_i\Big(\frac{\bar{z}-z_{\rm max}/2}{z_{\rm
max}/2}\Big)\Big] \ee The galaxy distribution of the $i$th bin is,
\be\label{gal_bin} N_i(z)\propto{N(z)}\Big[{\rm
erfc}\Big(\frac{z_{\rm i-1}-z}{\sqrt{2}\sigma(z)}\Big)- {\rm
erfc}\Big(\frac{z_{\rm i}-z}{\sqrt{2}\sigma(z)}\Big)\Big] \ee Thus
if the overall distribution is biased by $g$, all the redshift bins
are biased accordingly, as shown in Fig~\ref{fig:chb} (We show the
galaxy distributions of LSST for an example).

In our calculations we account for this effect by marginalising over $30$ Chebyshev
coefficients.\\

\begin{figure}[htp]
\hspace*{-0.8cm}\includegraphics[width=1.2\columnwidth]{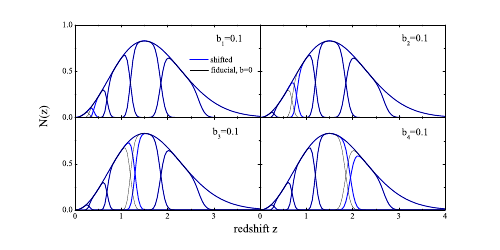}
\caption{Redshift error - shift: uncertainty in measuring the centroids of the redshift
bins (non-vanishing $z$-bias), quantified by $b_i$ coefficients at redshift bin $i$.} \label{fig:bias}
\end{figure}

{\it \underline{Redshift-bin Centroids Uncertainty:}}
To be general, we also consider the possible degradation if there
exists some uncertainty in measuring the centroids of the redshift
bins, i.e. the so-called non-vanishing $z$-bias
$b_i{\equiv}z_i-\bar{z}_i\neq0$. The $z$-bias basically `shifts' the
centroids of the bins. See this effect illustrated in
Fig~\ref{fig:bias}. In the calculation, we assign one $z$-bias
parameter for each bin, then marginalize over them.

Note that the centroid shifts do not capture the
catastrophic errors where a smaller fraction of redshifts are
completely mis-estimated and reside in a separate island in the
$z-\bar{z}$ plane~\cite{Huterer:2005ez,Bernstein:2009bq}.\\

{\it \underline{z-scatter:}}
We assume that $\sigma(z)=\sigma_{0}(1+z)$, and we choose
$\bar{\sigma}_{0}=0.03 (0.05)$ for fiducial model for LSST (DES).
But if $\sigma_{0}$ is not perfectly measured, i.e., the $z$-scatter
is not zero, $\Delta\equiv\sigma_0-\bar{\sigma}_0\neq0$, there will
be further degradation. In the calculation, we assign one $z$-scatter
parameter for each bin, then marginalize over. As shown
in~\cite{Huterer:2005ez}, this effect is sub-dominant.
\begin{figure}[htp]
\hspace*{-0.8cm}\includegraphics[width=1.2\columnwidth]{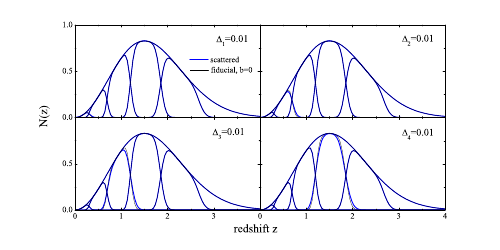}
\caption{Redshift error - scatter: uncertainties in the
photo-$z$ error at redshift $z$ modeled as $\sigma(z)=\sigma_{0}(1+z)$,
where $\Delta_i$ is $\sigma_0-\bar{\sigma}_0$ at redshift bin $i$.} 
\label{fig:scatter}
\end{figure}

\subsubsection{Additive errors}

Additive errors are present for both galaxy counts and lensing shear
measurements, and they are generated, for example, by the anisotropy
of the point spread function (PSF). Following \cite{Huterer:2005ez}
and \cite{Zhan:2008jh}, we parametrize the additive errors as
\be\label{add} (C_\ell^{\rm XY})_{ij}=\delta_{\rm XY}{\rho}A_i^{\rm
X}A_j^{\rm Y}\Big(\frac{\ell}{\ell_{\ast}^{\rm X}}\Big)^\gamma \ee and
choose $\rho=1, \gamma=0$. The fiducial values of the $A$'s are,
$(A^{\rm g})^2=10^{-8}, (A^{\gamma})^2=10^{-9}$ \cite{Zhan:2008jh}.

\subsubsection{Multiplicative errors}
Multiplicative errors in measuring shear can be introduced by
various sources. For example, a circular PSF of finite size is
convolved with the true image of the galaxy to produce the observed
image, and in the process it produces a multiplicative
error~\cite{Huterer:2005ez}. \be\label{mul}
(\widetilde{C}_{\ell}^\gamma)_{ij}=
(C_{\ell}^\gamma)_{ij}[1+f_i+f_j]\ee We choose $f=0$ as fiducial.

\end{document}